\documentclass[manuscript,screen]{acmart}
\usepackage{amsmath,amsfonts}
\usepackage{algorithmic}
\usepackage{algorithm}
\usepackage{array}
\usepackage[caption=false,font=footnotesize,labelfont=rm,textfont=rm]{subfig}
\usepackage{textcomp}
\usepackage{stfloats}
\usepackage{url}
\usepackage{verbatim}
\usepackage{graphicx}
\usepackage{listings}
\usepackage{xcolor} 
\usepackage{caption}
\usepackage{xspace}
\usepackage{makecell,multirow}
\usepackage{natbib}
\usepackage{enumitem}

\usepackage{tcolorbox}
\newcommand{\tool}{\textsc{BolaZ}\xspace}

\DeclareCaptionFormat{mylst}{\hrule#1#2#3}
\usepackage{orcidlink}

\AtBeginDocument{%
  }

\setcopyright{acmlicensed}
\copyrightyear{2025}
\acmYear{2025}
\acmDOI{XXXXXXX.XXXXXXX}
\acmConference{ACM Transactions on Software Engineering and Methodology}
\acmISBN{978-1-4503-XXXX-X/2018/06}

\begin{document}

\title{Rethinking Broken Object Level Authorization Attacks Under Zero Trust Principle}

\author{Anbin Wu}
\email{wuanbin@tju.edu.cn}
\orcid{0009-0003-6977-2252}
\author{Zhiyong Feng}
\email{zyfeng@tju.edu.cn}
\orcid{0000-0001-8158-7453}
\affiliation{%
  \institution{The College of Intelligence and Computing, Tianjin University}
  \country{CHINA}
}
\author{Ruitao Feng}
\orcid{0000-0001-9080-6865}
\affiliation{%
  \institution{Southern Cross University}
  \country{Australia}}
\email{ruitao.feng@scu.edu.au}
\authornote{Corresponding author}

\author{Zhenchang Xing}
\orcid{0000-0001-7663-1421}
\affiliation{%
  \institution{CSIRO’s Data61}
  \country{Australia}}
\email{zhenchang.xing@anu.edu.au}

\author{Yang Liu}
\orcid{0000-0001-7300-9215}
\affiliation{%
  \institution{School of Computer Science and Engineering, Nanyang Technological University}
  \country{Singapore}}
\email{yangliu@ntu.edu.sg}

\renewcommand{\shortauthors}{Anbin Wu et al.}

\begin{abstract}
  RESTful APIs facilitate data exchange between applications, but they also expose sensitive resources to potential exploitation. Broken Object Level Authorization (BOLA) is the top vulnerability in the OWASP API Security Top 10, exemplifies a critical access control flaw where attackers manipulate API parameters to gain unauthorized access. To address this, we propose \tool, a defense framework grounded in zero trust principles. \tool analyzes the data flow of resource IDs, pinpointing BOLA attack injection points and determining the associated authorization intervals to prevent horizontal privilege escalation. Our approach leverages static taint tracking to categorize APIs into producers and consumers based on how they handle resource IDs. By mapping the propagation paths of resource IDs, \tool captures the context in which these IDs are produced and consumed, allowing for precise identification of authorization boundaries. Unlike defense methods based on common authorization models, \tool is the first authorization-guided method that adapts defense rules based on the system’s best-practice authorization logic. We validate \tool through empirical research on 10 GitHub projects. The results demonstrate \tool's effectiveness in defending against vulnerabilities collected from CVE and discovering 35 new BOLA vulnerabilities in the wild, demonstrating its practicality in real-world deployments.
\end{abstract}



\maketitle

\section{Introduction}
RESTful APIs have become the standard for accessing web-oriented resources, enabling users to initiate operational requests through HTTP methods, paths, and parameters. However, the parameters of RESTful APIs are user-controlled, hence, inherently untrusted. Attackers can exploit this by tampering with the \textit{\textbf{resource ID}} parameter to access sensitive data of other users without authorization, leading to a \textbf{Broken Object Level Authorization (BOLA) attack} \cite{idris_development_2021}. Resource ID is the unique identifier of resources in REST API that are used to operate (INSERT, DELETE, UPDATE, READ) resources, that is, \textbf{BOLA attack injection point}. The vulnerability arises from over-reliance on the user-supplied resource ID, without enforcing proper access control mechanisms.

The BOLA vulnerability is an access control vulnerability, and developers need to perform some security checks before users operate sensitive information to defend against the vulnerability. BOLA vulnerabilities are derived from the lack of object-level authorization checking \cite{huang_detecting_2024}. To accurately detect BOLA vulnerabilities, it is necessary to understand the access control policy of resources. Through the study of the existing work, there are two main ways to obtain access control policies. 1). Manually provide authorization policies rules \cite{dalton_nemesis_2009}, \cite{muthukumaran_flowwatcher_2015}, \cite{bocic_finding_2016}. This method relies on manually labeling authorization rules for each resource operation, which is tedious and prone to errors \cite{ghorbanzadeh_anovul_2020}. 2). Deduce the authorization model of resources through source code \cite{son_rolecast_2011}, \cite{monshizadeh_mace_2014}, \cite{huang_detecting_2024},  \cite{near_finding_2016}. This type of method first artificially summarizes the application's resource authorization model, and infers which authorization model the resource belongs to by analyzing the source code. However, this type of method is based on the authorization model of human summary. When the system authorization model is inconsistent with the authorization model of human summary, it will lead to false positives. The authorization model's static nature and the unknown system logic's dynamic nature are contradictory.
\begin{figure}[htbp]
	\centering
	\includegraphics[scale=0.4]{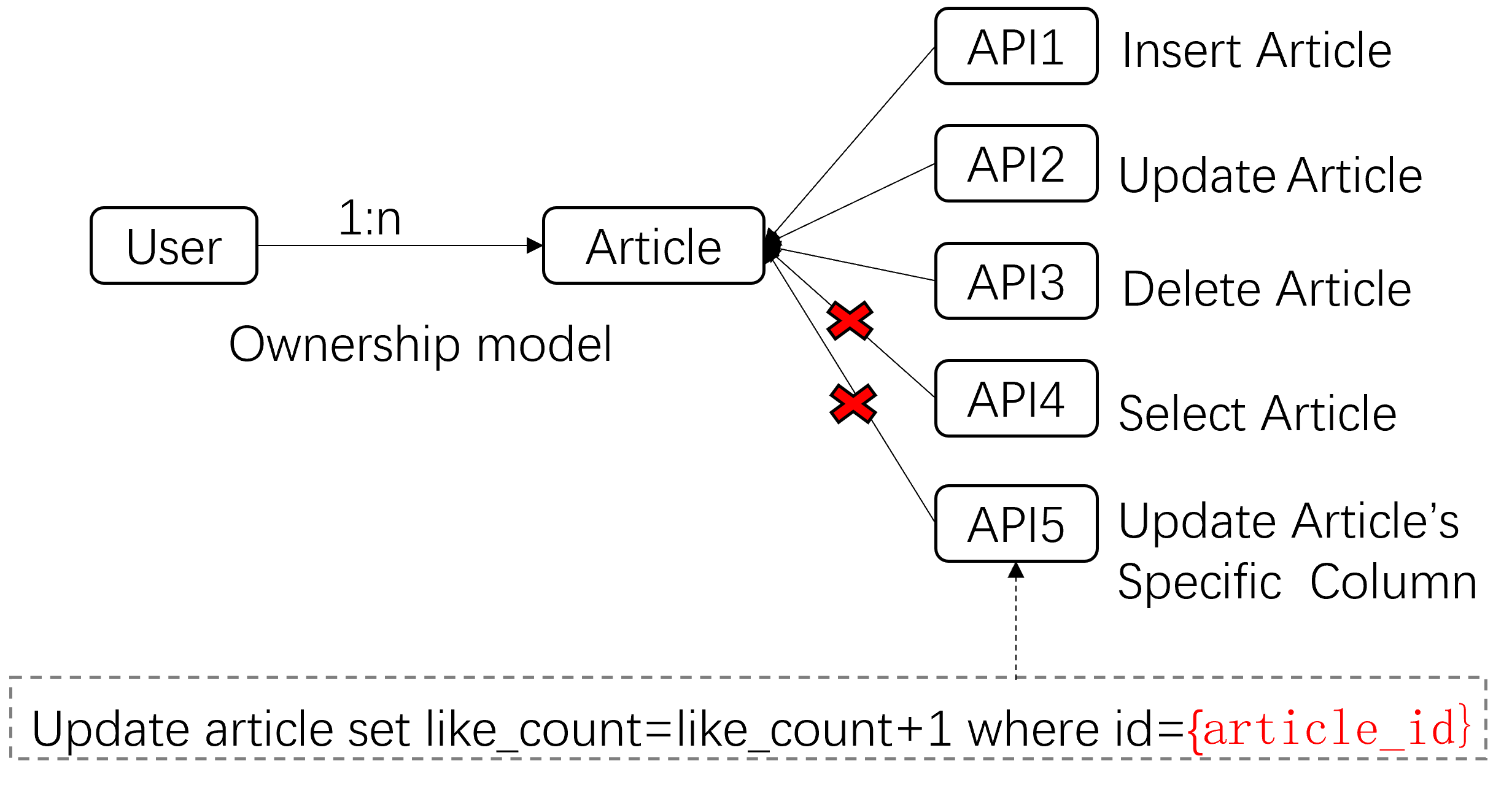}
	\caption{BOLARAY authorization model}
	\label{fig_0}
\end{figure}

BOLARAY (CCS’24, Oct) \cite{huang_detecting_2024} is the state-of-the-art (SOTA) method for detecting BOLA attacks. By analyzing real-world BOLA vulnerabilities in open-source applications, BOLARAY identifies four common object-level authorization models. However, these models, being manually summarized, lack adaptability to the diverse and evolving nature of modern web applications. As illustrated in Fig.\ref{fig_0}, BOLARAY classifies the \emph{Article} table under the ownership model, where an article can only be updated or deleted by its owner. This model fails to support SELECT operations. For instance, API5 in Fig.\ref{fig_0} involves an UPDATE statement that allows users to add likes to articles. However, BOLARAY incorrectly assumes that articles can only be updated by their owners. This rigid interpretation overlooks the actual propagation patterns of resource identifiers in web applications \cite{akamai_owasps-top-10-api-security-risks_2024}, leading to semantic mismatches. As a result, such out-of-context authorization logic limits BOLARAY’s ability to comprehensively detect BOLA vulnerabilities.


We revisit the essence of BOLA attacks—horizontal privilege escalation and reconceptualize their defense as a problem of minimally scoped authorization. Specifically, defending against BOLA attacks involves isolating the injection point (i.e., the resource ID) within the narrowest permission boundary that aligns with the application's logic. The resource ID parameter functions dually as an attack surface vulnerable to horizontal privilege escalation \cite{sengupta_survey_2020}, and as a protection surface responsible for enforcing least privilege access control \cite{campbell_beyond_2020}. Effective mitigation requires accurately determining the authorization interval for each resource ID, ensuring users can only access or manipulate resources within their minimum necessary permissions \cite{jero_practical_2021}.

    
Zero Trust, grounded in the principle of ``never trust, always verify" \cite{samaniego_zero-trust_2018}, emphasizes fine-grained, identity-based access control to mitigate risks associated with unauthorized lateral movement. \textit{\textbf{Micro-segmentation (MSG)}}, a crucial component of zero trust, creates a security closed-loop by considering multiple factors—such as user roles, processes, and access contexts—to define the minimal security boundary for resource IDs based on API context data flow. This not only aligns with the API's authorization logic but also enforces the principle of minimum permissions, making MSG an effective strategy for defending against BOLA attacks.

This paper introduces \tool \footnote{Code availability: https://anonymous.4open.science/r/bolaz-96AC}, a novel framework designed to defend against BOLA attacks from two perspectives: attack surface discovery and protection surface authorization. \tool leverages an attacker's viewpoint to analyze BOLA attack data flows, accurately \textbf{{identifying the BOLA attack injection points}}. \tool takes advantage of the context relationship between API parameters and resource ID to compensate for the shortcomings of identifying injection points based on fixed authorization models. Additionally, \tool divides resources into smaller logical intervals based on resource ID workflows, isolating the protection surface to the \textbf{{minimal authorization intervals}} defined by the system's authorization logic. This approach enables effective resource isolation across different users and business processes, overcoming the limitations of static authorization models.
	
Through empirical research on 10 GitHub projects verified to contain RESTful API usage, we first quantitatively evaluate \tool's effectiveness by assessing the precision and recall of attack injection point identification, authorization interval determination, and performance overhead. Our experimental selection criteria—GitHub stars, application type, and logic control type (detailed in Section~\ref{sec:MSG Rules Of Resource ID})—ensure a comprehensive coverage of various application scenarios. Next, we collect relevant vulnerabilities from the Common Vulnerabilities and Exposures (CVE) DB based on three BOLA attack modes (detailed in Section~\ref{sec:Threat model}) to demonstrate \tool's capability in detecting these vulnerabilities. Finally, we apply \tool to scan for potential BOLA vulnerabilities in 10 projects, evaluating it in real-world scenarios to confirm its practicality.
	
To the best of our knowledge, this is the first work to apply Micro-segmentation (MSG), which is derived from zero trust principles, to the discovery of attack surfaces and the refinement of data-level permissions in RESTful APIs. Among the 94 RESTful APIs analyzed in the selected GitHub projects, \tool achieved recall rates of 97\% for attack injection points and 87\% for authorization intervals, with only one false positive. The experimental results are promising; \tool successfully defended against known BOLA vulnerabilities in our benchmark and identified 35 new BOLA vulnerabilities in 10 real-world projects. 
    
This paper makes the following contributions:
	
	\begin{itemize}
		\item We propose a novel approach for \textbf{identifying BOLA attack injection points} by tracking the data flow of resource IDs. This method enhances detection by analyzing how resource IDs are used throughout the system.
		\item Our work is the first to apply static taint tracking technology to \textbf{infer authorization intervals (MSG intervals)} for BOLA attack injection points. This technique improves the granularity of access control by defining minimum security boundaries based on data flow.
		\item We introduce a method to \textbf{isolate attack injection points to the system's minimum security boundary}, leveraging the resource ID propagation mode. \tool is the \textbf{first authorization-guided} method that adapts defense rules based on the system's best-practice authorization logic.
	\end{itemize}

The rest of this paper is organized as follows. Section \ref{background} introduces background knowledge on RESTful APIs, BOLA attacks and Micro-Segmentation (MSG). Section \ref{problem_formulation} demonstrates how the problem of BOLA vulnerability detection is formulated at the methodological level. In Section \ref{approach}, the technical details of the proposed novel approach, \tool, are described. Section \ref{implementation} shows the details of the implementation. Section \ref{evaluation} describes the experiments and discusses the results. Section \ref{sec:approach_limitations} provides discussion and future work. Relevant literature is summarized in Section \ref{section:7}. Section \ref{section:8} concludes this work.

\section{Background}\label{background}
\subsection{RESTful APIs}
	RESTful APIs are APIs that follow the style of REST architecture \cite{fielding_architectural_2000}. RESTful APIs provide a unified interface for creating (C), reading (R), updating (U), and deleting (D) resources \cite{corradini_automated_2023}. RESTful APIs are a resource-oriented architecture. Users can identify resources through the URI and correspond to the CRUD operation of the resource through POST, GET, UPDATE, DELETE and PATCH.

	\subsection{BOLA Attack}
	BOLA is the number one vulnerability in OWASP API Security \cite{owasp_api_security_owasp_2023}. Let's take a concrete example to explain what the BOLA attack is. As shown in Fig.\ref{fig_1_1}, users using order details API (\emph{GET /api/order/\{orderid\}}) can only read order resources that they have created. Normal users (userA, userB) access the \emph{order} resource using their own created \emph{orderid} (\textbf{733,845}). Still, the attacker initiates access to userB's order resources by modifying the \emph{orderid} (\textbf{\textit{resource ID}}) parameter, which is the \textbf{injection point} of the BOLA attack. If the developer does not perform an effective resource ID permission check on the API, attackers can gain unauthorized access to other users' sensitive resources. The BOLA vulnerability opens the door for attackers to directly access resources, allowing them to \textbf{{bypass intended application workflows}} and gain unauthorized access to sensitive data.
 
	\subsection{Micro-Segmentation (MSG)}
\begin{figure}[htbp]
		\centering
		\subfloat[BOLA Attack]{\includegraphics[scale=0.4]{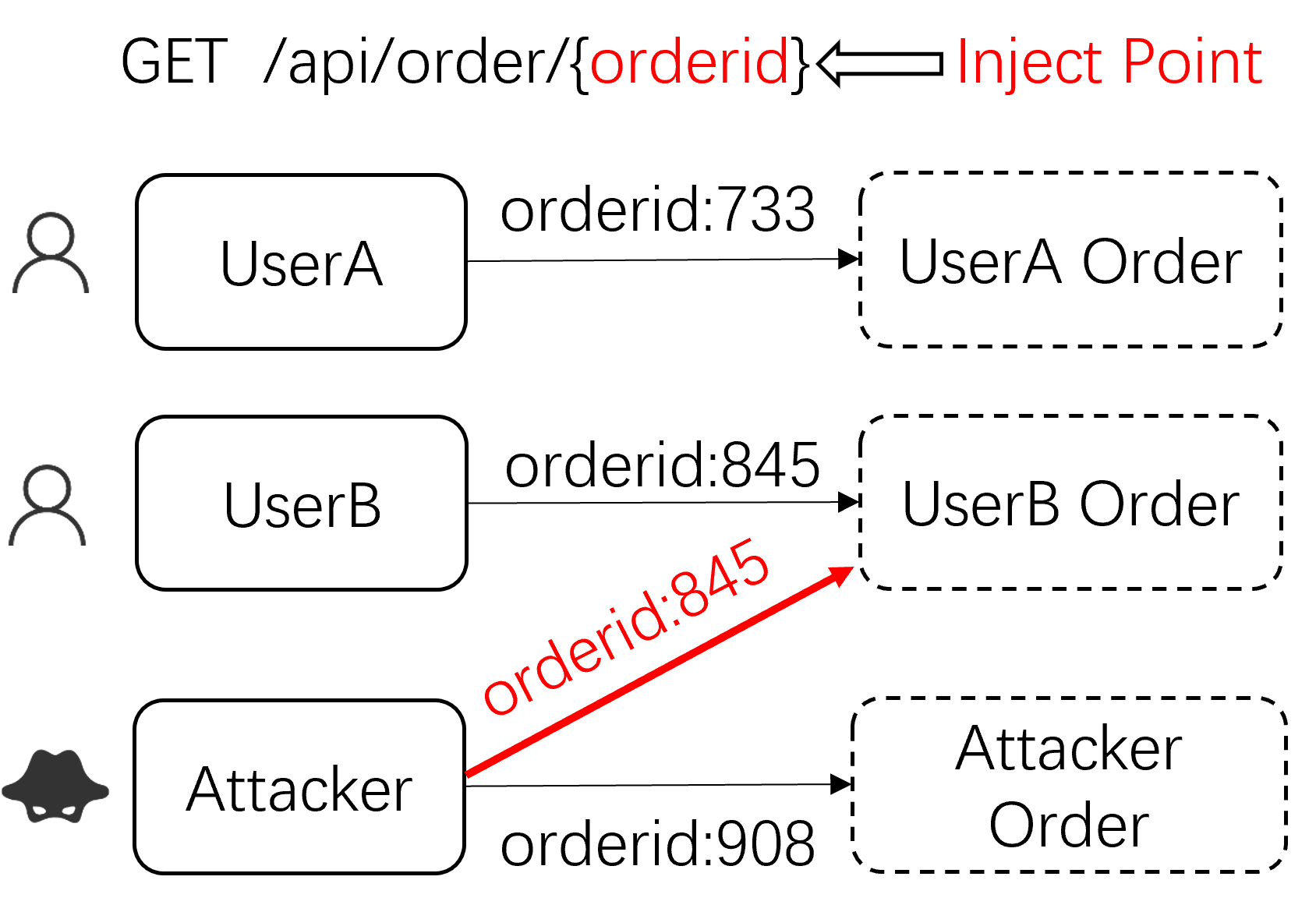}%
			\label{fig_1_1}}
		\hfil
		\subfloat[MSG]{\includegraphics[scale=0.4]{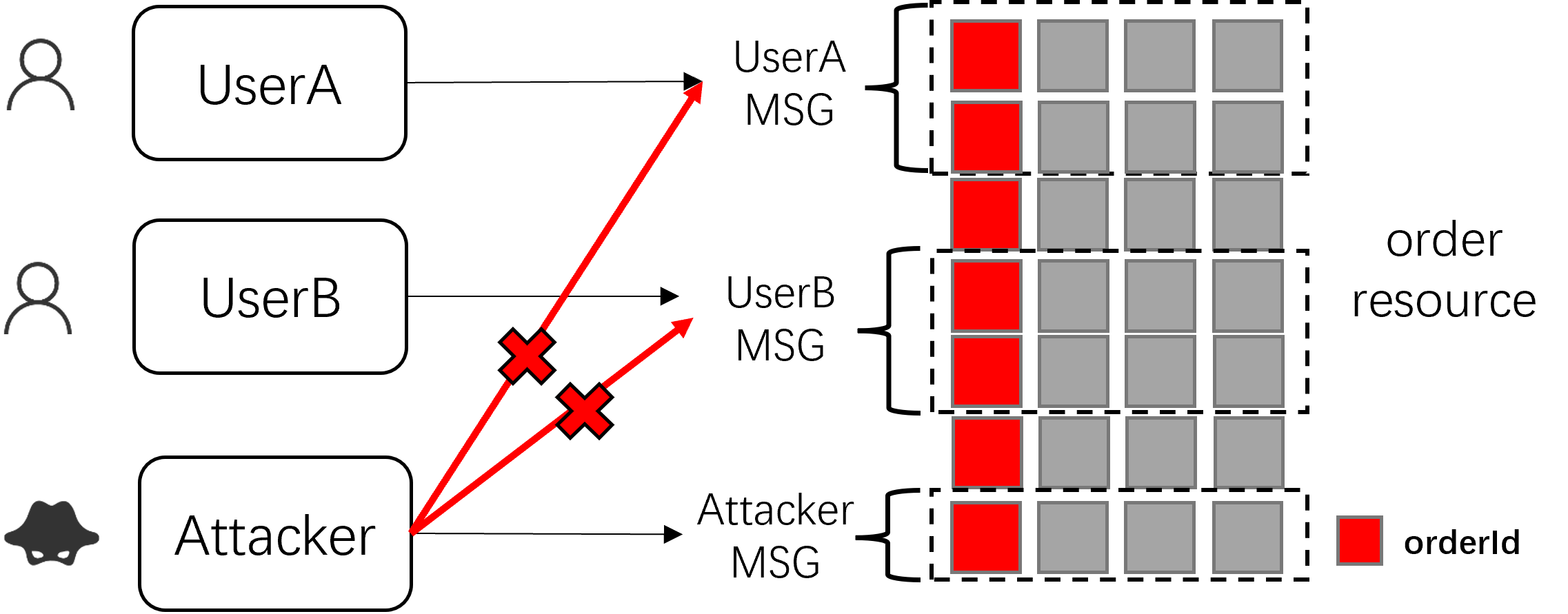}%
			\label{fig_1_2}}
		\caption{A BOLA attack on an order details API and its prevention by Micro-Segmentation (MSG)}
		\label{fig_1}
	\end{figure}

	MSG \cite{keeriyattil_microsegmentation_2019} follows the core principles of zero trust-minimum permissions, all users are untrusted, and user access to any resource needs to be authenticated. MSG divides resources into several small resource intervals based on service logic in a software-defined way, logically separating resources and restricting the movement of users within resources, enabling the most fine-grained access control of resources. As shown in Fig.\ref{fig_1_2}, according to the system logic, MSG divides the entire \emph{order} resource into small partitions according to user identity. MSG policies allow users to access only the \textit{orderid} created by themselves, and prevents the attacker's horizontal unauthorized access to other users' \textit{order} resources.
    
\section{Problem Formulation}\label{problem_formulation}

\subsection{Threat Model} \label{sec:Threat model}
	\subsubsection{Attacker's Capability} 
\begin{figure}[htbp]
			\centering
			\includegraphics[scale=0.6]{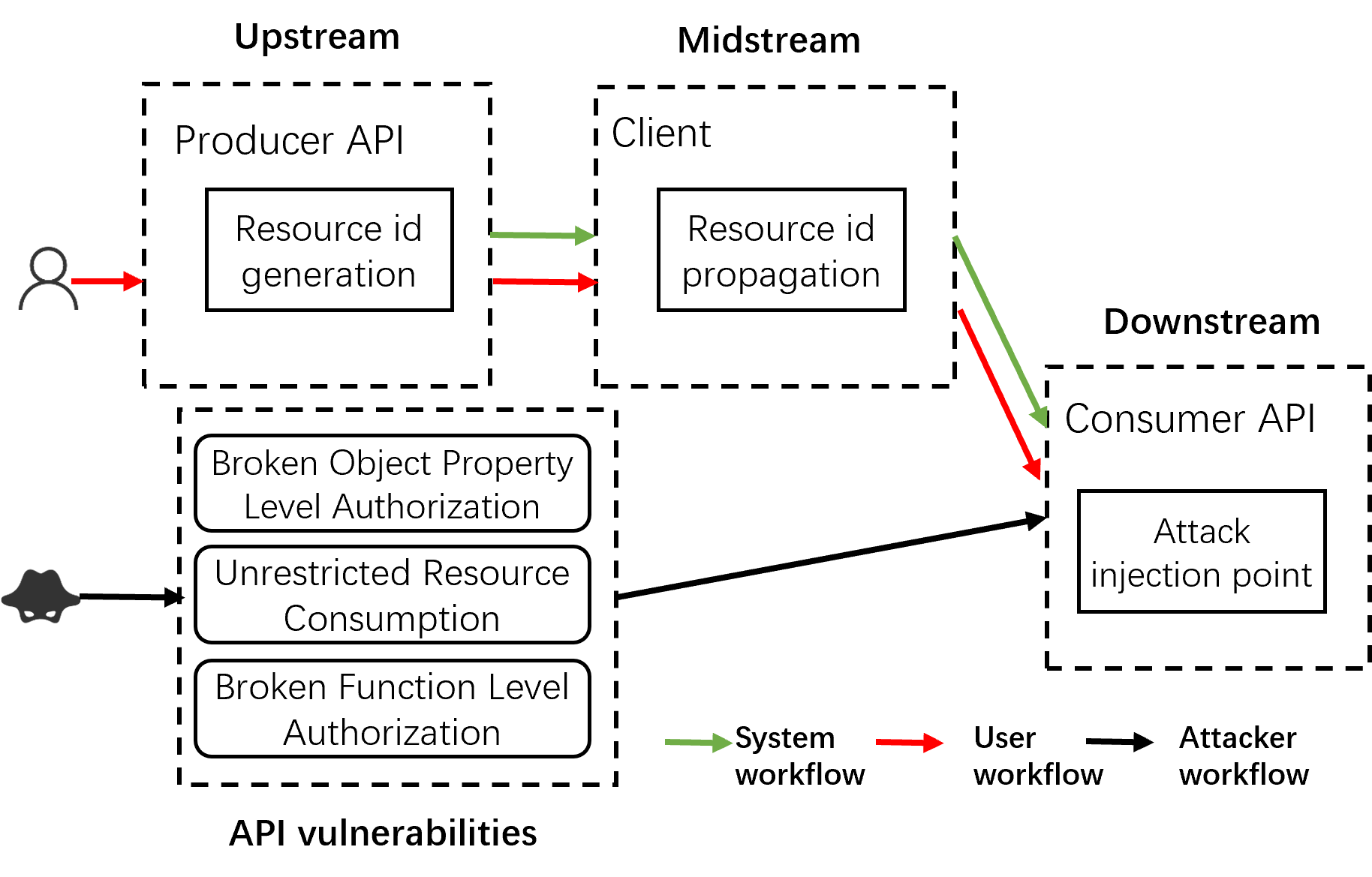}
			\caption{System, user, and attacker workflows}
			\label{fig_2}
	\end{figure}

	To effectively execute a Broken Object Level Authorization (BOLA) attack, attackers must obtain resource IDs they are not authorized to access. These IDs can be exposed through several API vulnerabilities, as shown in Fig.\ref{fig_2}. The primary vulnerabilities leading to BOLA attacks are Broken Object Property Level Authorization (BOPLA), Unrestricted Access to Sensitive Business Flows (UASBF), and Broken Function Level Authorization (BFLA) \cite{owasp_api_security_owasp_2023}.
	
	\subsubsection{Vulnerability Techniques} \label{sec:Technique}
	\begin{itemize}[left=0em,label={-}]
		\item {\textbf{BOPLA}}, ranked 3rd in the OWASP API Security Top 10, occurs when APIs unintentionally expose sensitive resource IDs in their responses. Attackers exploit this exposure by extracting these IDs from the API responses and using them to gain unauthorized access to resources.
		\item {\textbf{UASBF}}, ranked 4th, allows attackers to bypass restrictions and access APIs without proper authentication. When resource IDs are predictable or sequential, attackers can use automated scripts to enumerate these IDs, leading to potential BOLA attacks as they gain unauthorized access to various resources.
		\item {\textbf{BFLA}}, listed 5th, allows attackers to access APIs with elevated permissions without proper authorization checks. If an API fails to enforce authorization controls, attackers may exploit this to retrieve sensitive resource IDs belonging to other users.
	\end{itemize}

	
	
	
	\subsection{Terminology}
\begin{figure}[htbp]
    \centering
			\includegraphics[scale=0.5]{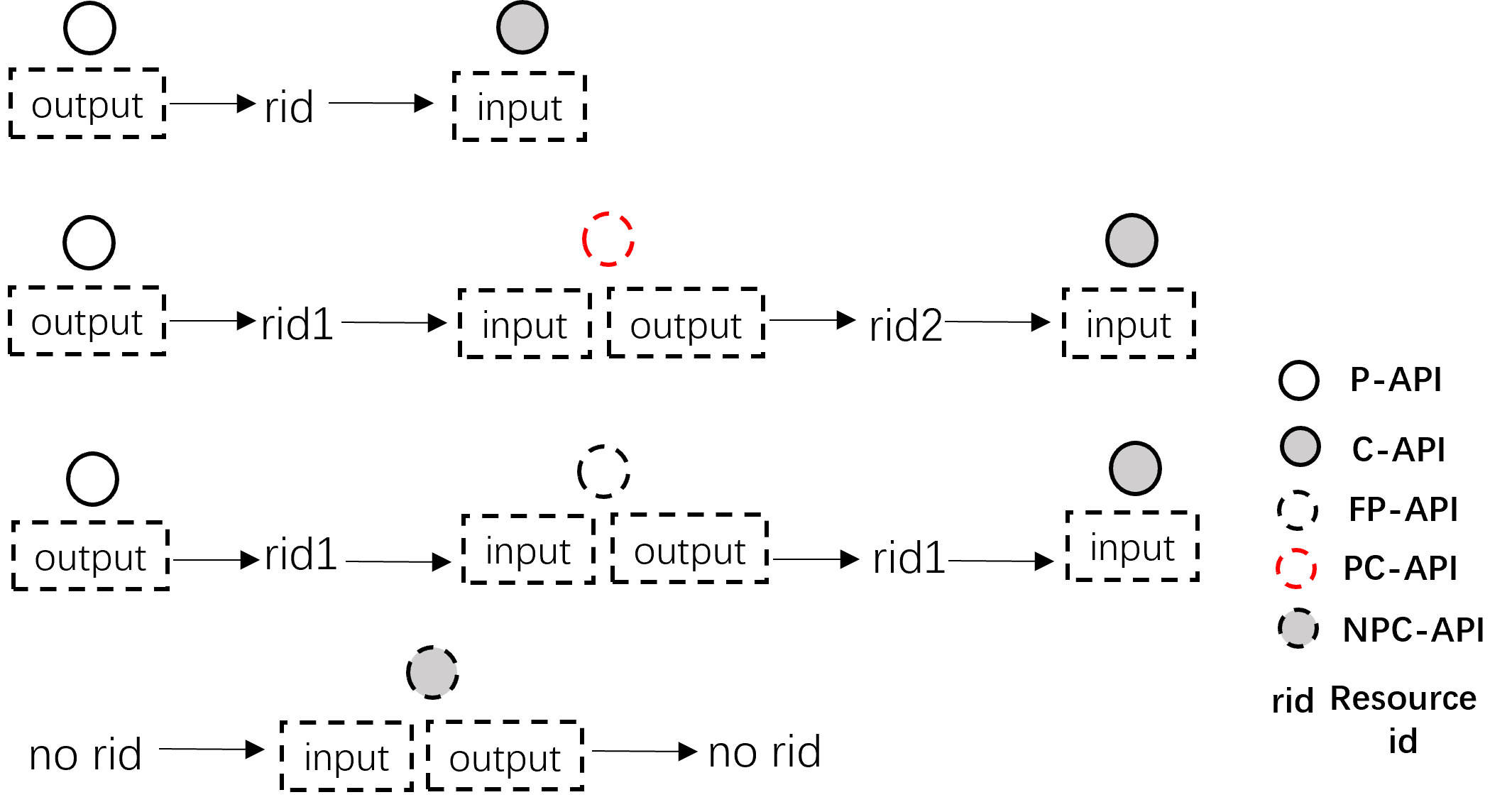}
			\caption{Classification of APIs based on rid production and consumption}
			\label{fig_3}
    \end{figure}

	We classify APIs according to the production and consumption of resource IDs, as shown in Fig.\ref{fig_3}.
	\begin{itemize}[left=0em,label={-}]
		\item \textbf{\emph{Producer API (P-API).}} P-APIs can return the resource ID.
		\item \textbf{\emph{Consumer API (C-API).}} C-APIs consume the resource ID through parameters.
		\item \textbf{\emph{False Producer API (FP-API).}} FP-APIs consume and return the same resource ID, which is essentially C-API.
		\item \textbf{\emph{Producer and consumer API (PC-API).}} PC-APIs consume some resource IDs and produce other types of resource IDs.
		\item \textbf{\emph{Non-producer and non-consumer API (NPC-API).}} NPC-APIs neither produce nor consume resource IDs.
	\end{itemize}

	\subsection{Problem Definition} \label{sec:ProblemDefinition}
	Defending against BOLA attacks fundamentally involves confining the resource ID accessed by attackers within the smallest necessary permission scope. As illustrated in Fig.\ref{fig_2}, the resource ID (attack injection point) is not arbitrarily provided by the user; it is generated upstream in the system workflow and then propagated downstream. Then the resource ID generated upstream of the data stream is the accessible interval of the resource ID of the downstream consumer API. Consequently, from the perspective of the resource ID's data flow, the problem of defending against BOLA attacks can be decomposed into three sub-problems.
	
	\noindent\textbf{{Problem 1: P-API identification and discovery of resource ID generation rules.}} The P-API is located upstream in the resource ID data stream, which generates a set of resource IDs according to the system's logical constraints. Therefore, the first sub-problem is to identify the P-API and extract the rules governing resource ID generation.
	
	\emph{Formulation:} Given an API $P$, determine whether it produces the resource ID $rid$ and the logic rules $R$ of creating the resource ID.
	\begin{equation}
		P \in API,Rule=\{R_1,...,R_n\},R_i=\{rid_1,...rid_n\}
	\end{equation}
	
	\noindent\textbf{{Problem 2: C-API identification and discovery of BOLA attack injection points.}} The C-API is downstream in the resource ID data stream, utilizing the resource ID propagated by the P-API to perform resource operations. Therefore, the second sub-problem is to identify the C-API and find the BOLA attack injection points in the C-API.
	
	\emph{Formulation:} Given an API $C$, determine whether it consumes the resource ID $rid$ and which parameters $parm$ are associated with the resource ID.
	\begin{equation}
		C \in API, parm_i \in C,C_{relation}=\{(parm_i,rid_i)\}
	\end{equation}
	
	\noindent\textbf{{Problem 3: Data flow association between P-API and C-API.}} P-API and C-API serve as the resource ID data flow's starting and ending nodes. Analyzing the data flow relationship between P-API and C-API can determine the authorization interval of the BOLA attack injection points.
	
	\emph{Formulation:} Given a P-API, P, and a C-API, C, determine whether there is a data flow context between P and C.
	\begin{equation}
		\begin{aligned}
			P \in P-API, C \in C-API, API_{relation}=(P,C)
		\end{aligned}
	\end{equation}
	
	\subsection{Problem Scope}
	There are many ways to store and extract resources in RESTful APIs. We define the problem scope to reduce the divergence of the research process and propose an extensible method based on this.
	
	\subsubsection{Resource Storage Based on Database}
	There are many forms of resource storage in RESTful APIs, such as databases, Hadoop, and Spark. The database is the main carrier for resource storage. Therefore, \tool considers the database as a resource storage carrier for BOLA attack detection, the same as BOLARAY~\cite{huang_detecting_2024}.
	
	\subsubsection{MSG Intervals = SELECT Statements} \label{sec:MSG Rules Of Resource ID}
     \tool uses the database as the resource storage carrier, so SELECT statements are the core way for P-API to obtain resource IDs, and SELECT statements support most of the data filtering functions. Server-side codes rarely filter resource IDs received by SELECT statements. Therefore, \tool uses SELECT statements of P-APIs as MSG intervals of resource IDs. The \textbf{\emph{primary and foreign keys}} obtained by SELECT statements are \textbf{\emph{resource IDs}}. 
	
	MSG intervals are not only determined by SELECT statements because there are two ways for users to delete and update resources: 1). \textbf{Server logic control}. P-APIs generate resources that users have permission to delete and modify, and users can delete and modify all resources returned by P-APIs on the client. 2). \textbf{Client logic control}. P-APIs return all resources to clients, and front-end codes determine whether current users have permission to use the resource ID to delete or modify the resource.
    
   \subsubsection{Propagation invariance of resource IDs}
   \label{sec:Propagation invariance of resource IDs}
   By analyzing nearly a hundred open source projects, we find that the resource ID, as the unique identifier of the resource, has assignment and replication operations during the propagation process, and there are few modification and deletion operations (only one case is found in nearly 1,000 APIs). Resource IDs remain invariant over the entire data stream of production, propagation and consumption. Moreover, in API automated testing, the resource ID is also regarded as immutable, e.g., RestTestGen and RESTler.

\section{Approach}\label{approach}
\begin{figure*}[htbp]
		\centering
		\includegraphics[scale=0.55]{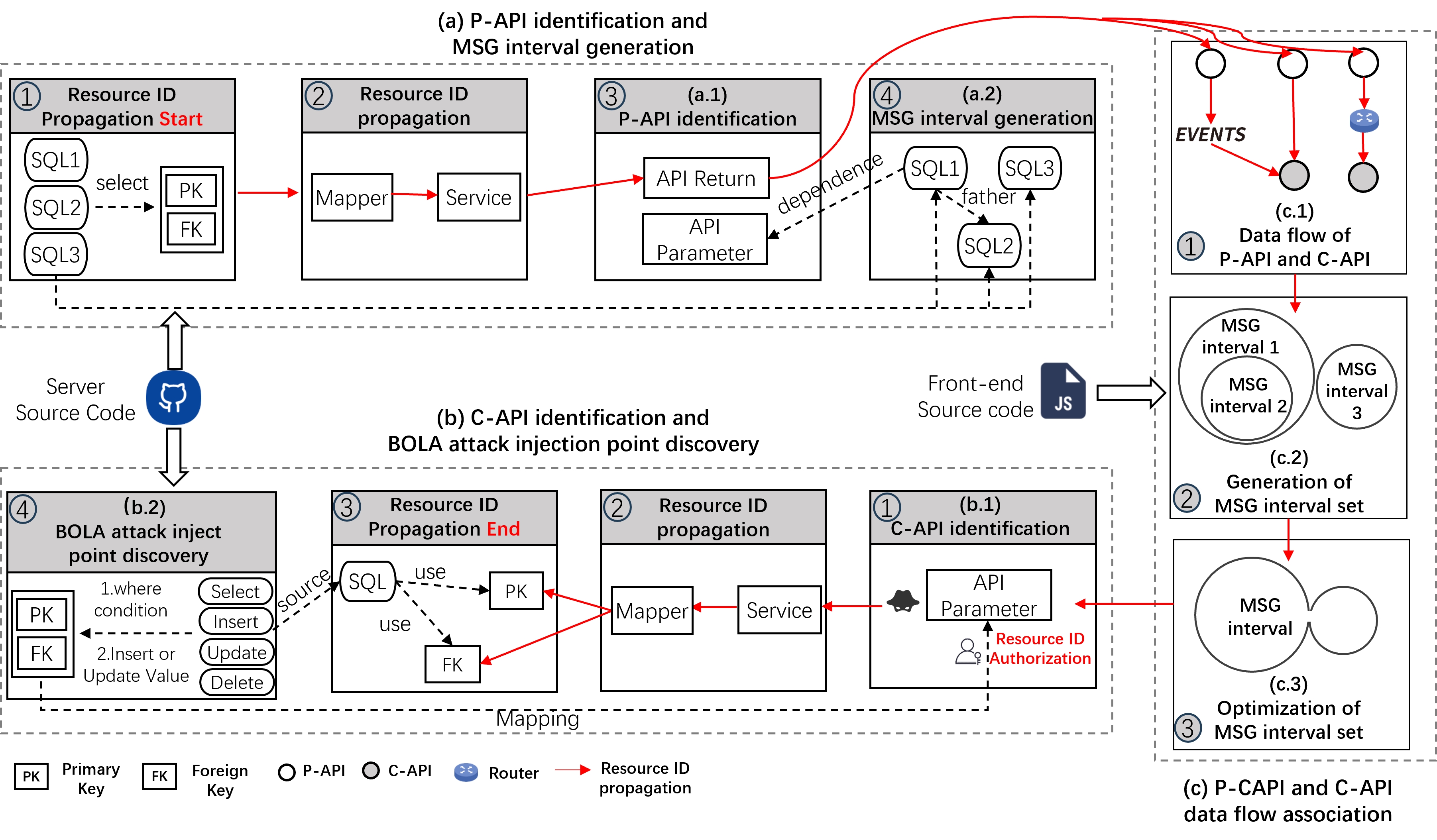}
		\caption{The overview of \tool}
		\label{fig_4}
	\end{figure*}

In this section, we give an overview of the \tool framework and propose three modules to solve our three sub-problems as described in Fig.\ref{fig_4}.
	
	\subsection{Overview}\label{sec:Overview}
	
	\textbf{\emph{(a) P-API identification and MSG interval generation.}} \normalsize{\textcircled{\scriptsize{1}}} \tool analyzes which SELECT statements generate resource IDs (primary and foreign keys) from server-side source code. 
	\normalsize{\textcircled{\scriptsize{2}}} Resource IDs generated by SELECT statements are propagated in the Mapper layer (database processing layer) and the Service layer (application logic processing layer). 
	\normalsize{\textcircled{\scriptsize{3}}} Resource IDs are propagated to the return value of the API. \tool determines if the API is P-API. 
	\normalsize{\textcircled{\scriptsize{4}}} \tool analyzes the dependencies between P-API parameters and SELECT statements, explores the parent-child relationship between SELECT statements, and generates MSG intervals of resource ID.
	
	\noindent\textbf{(b) \emph{C-API identification and BOLA attack injection point discovery.}} \normalsize{\textcircled{\scriptsize{1}}} \tool tracks the data flow of API parameters in the server source code. 
	\normalsize{\textcircled{\scriptsize{2}}} API parameters are propagated at the Service and Mapper layers.
	\normalsize{\textcircled{\scriptsize{3}}} \tool only retains the API parameter data flow propagating to the SQL statement's primary or foreign key (resource ID). 
	\normalsize{\textcircled{\scriptsize{4}}} \tool uses the mapping relationship between API parameters and primary and foreign keys to identify C-API and finds the injection point of the BOLA attack.
	
	\noindent\textbf{(c) \emph{P-API and C-API data flow association.}} \normalsize{\textcircled{\scriptsize{1}}} \tool explores data flows of resource IDs in front-end code from P-API to C-API and innovatively obtains the relationship between P-API and C-API. \normalsize{\textcircled{\scriptsize{2}}} MSG intervals of multiple P-APIs provide MSG intervals for attack injection points of C-APIs and generate the MSG interval set. \normalsize{\textcircled{\scriptsize{3}}} \tool improves the performance of resource ID authorization checking by optimizing the MSG interval set. 
	
	\subsection{P-API Identification and MSG Interval Generation}
	The P-API is the upstream of the resource ID propagation and generates the resource ID's MSG interval. Therefore, \tool's first step is to identify the P-API and get the MSG interval of the resource ID.

\begin{figure}[htbp]
	\centering
	\includegraphics[scale=0.6]{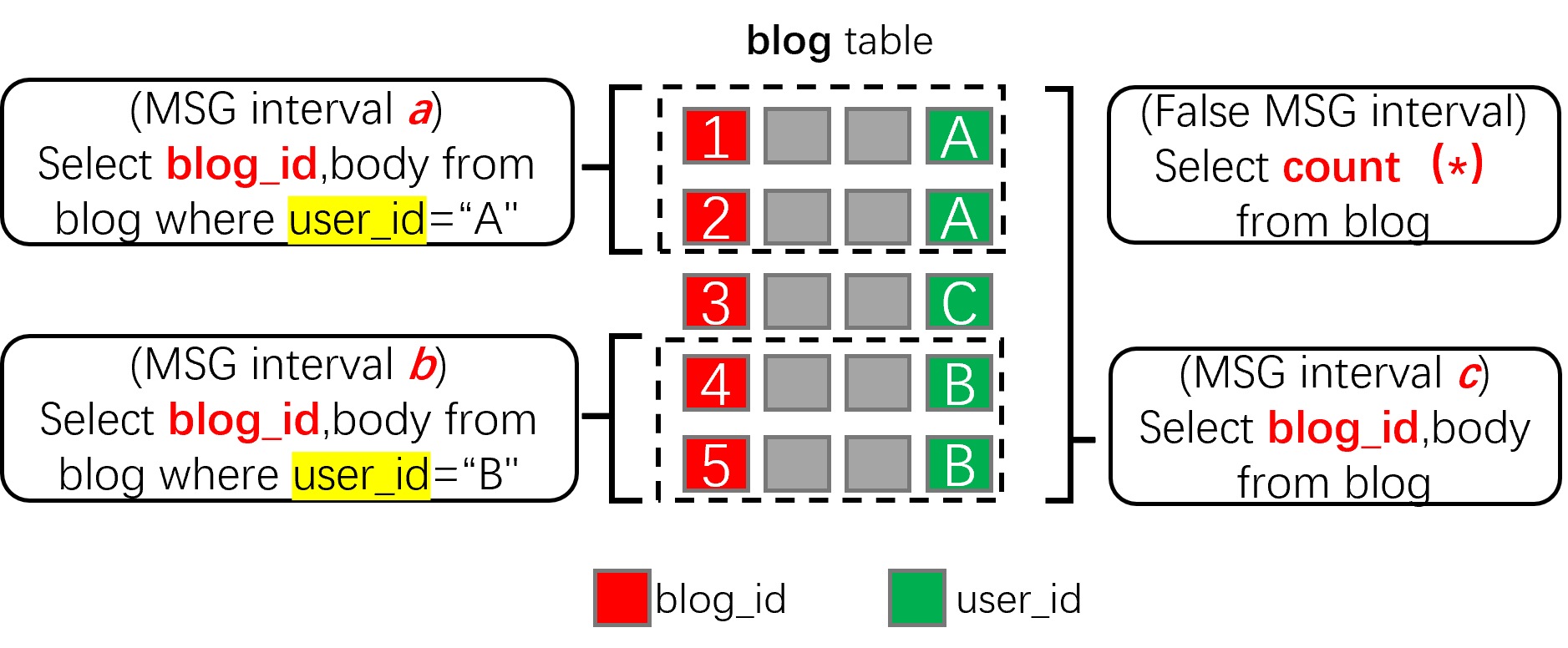}
	\caption{MSG intervals}
	\label{fig_4_0}
    \end{figure}

    \tool argues that SELECT statements represent MSG intervals. As shown in Fig.\ref{fig_4_0}, the three SELECT statements produce MSG intervals for the resource ID(\emph{blog\_id}) of \emph{blog}, respectively. The \emph{MSG interval a} contains the \emph{blog\_id} created by user A himself.  The \emph{MSG interval b} contains the \emph{blog\_id} created by user b himself. The \emph{MSG interval c} contains all \emph{blog\_id}. The false MSG interval only reads the amount of data in the \emph{blog} table and does not generate resource ID.
    
	\subsubsection{P-API Identification}
	Usually, the GET type API returns the resource ID \cite{zhang_resource_2021}. \tool takes the GET type API as the starting point and uses the taint tracking technology to obtain SELECT statements associated with the return value in the API.
    
\begin{figure}[htbp]
    \centering
    \subfloat[P-API identification]{
    		\includegraphics[scale=0.48]{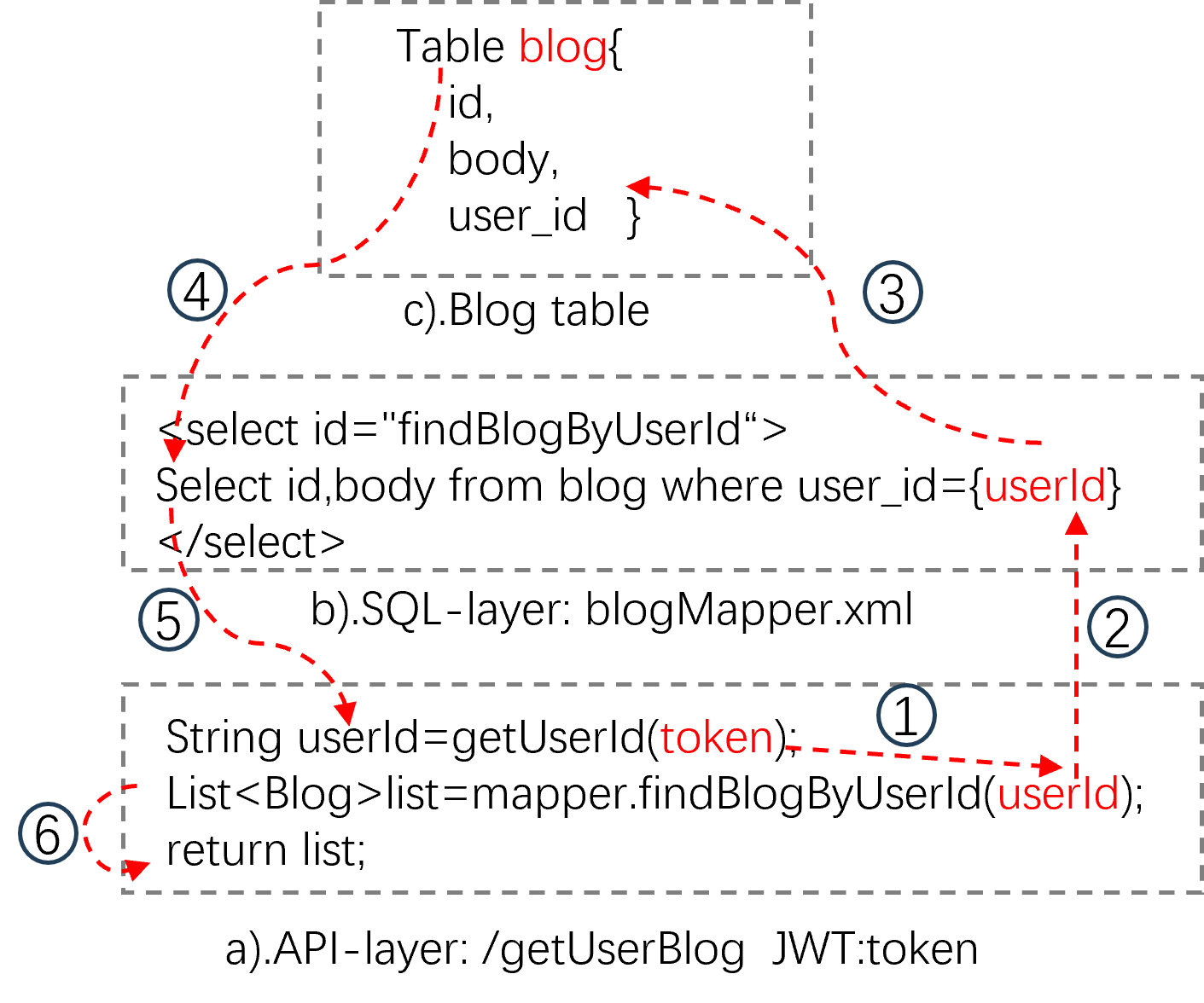}\label{fig_4_1}}
            \quad
    \subfloat[C-API identification]{
    		\includegraphics[scale=0.48]{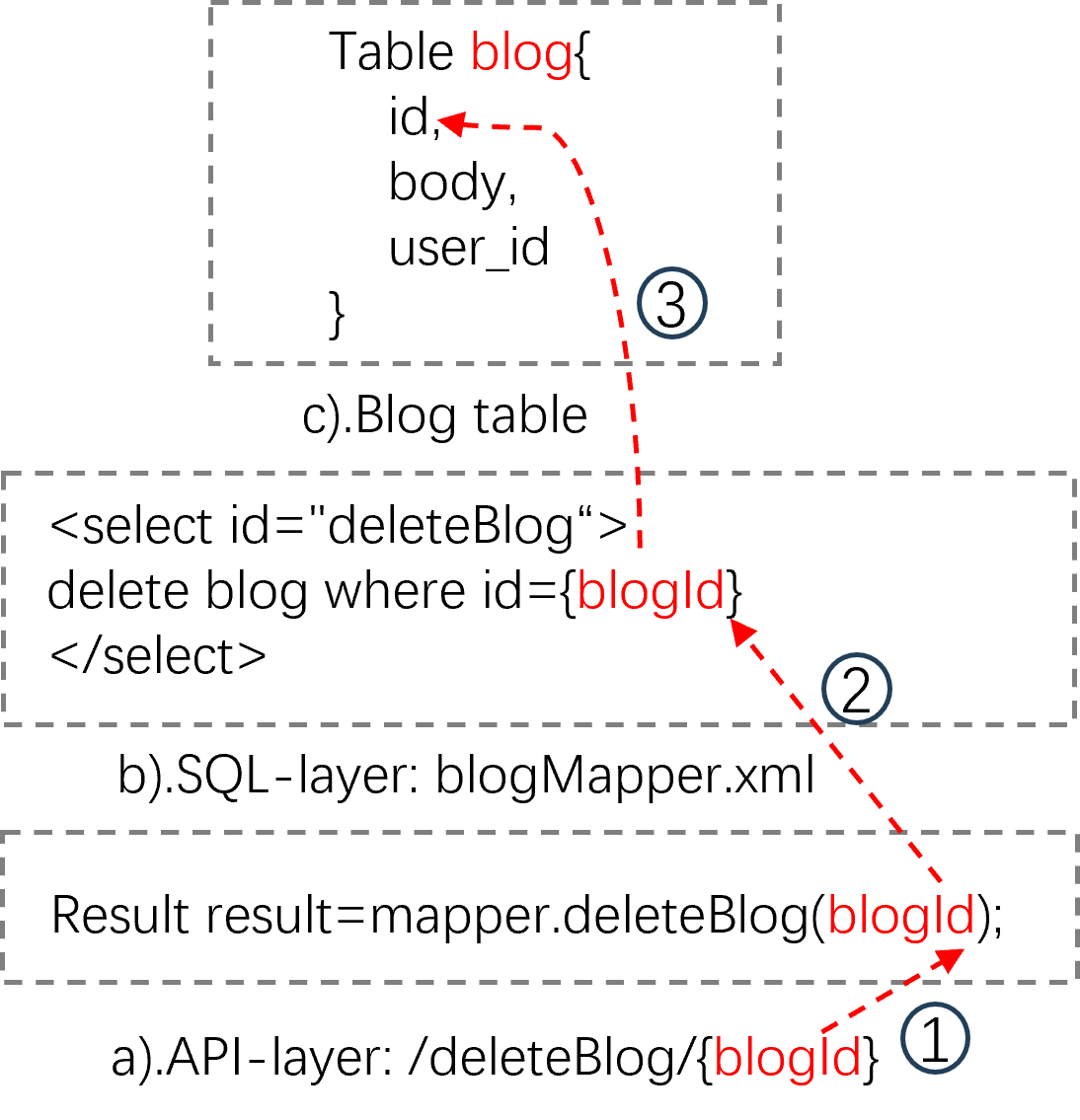}\label{fig_7_1}}
            \quad
    \subfloat[P-API and C-API data flow association]{
    		\includegraphics[scale=0.48]{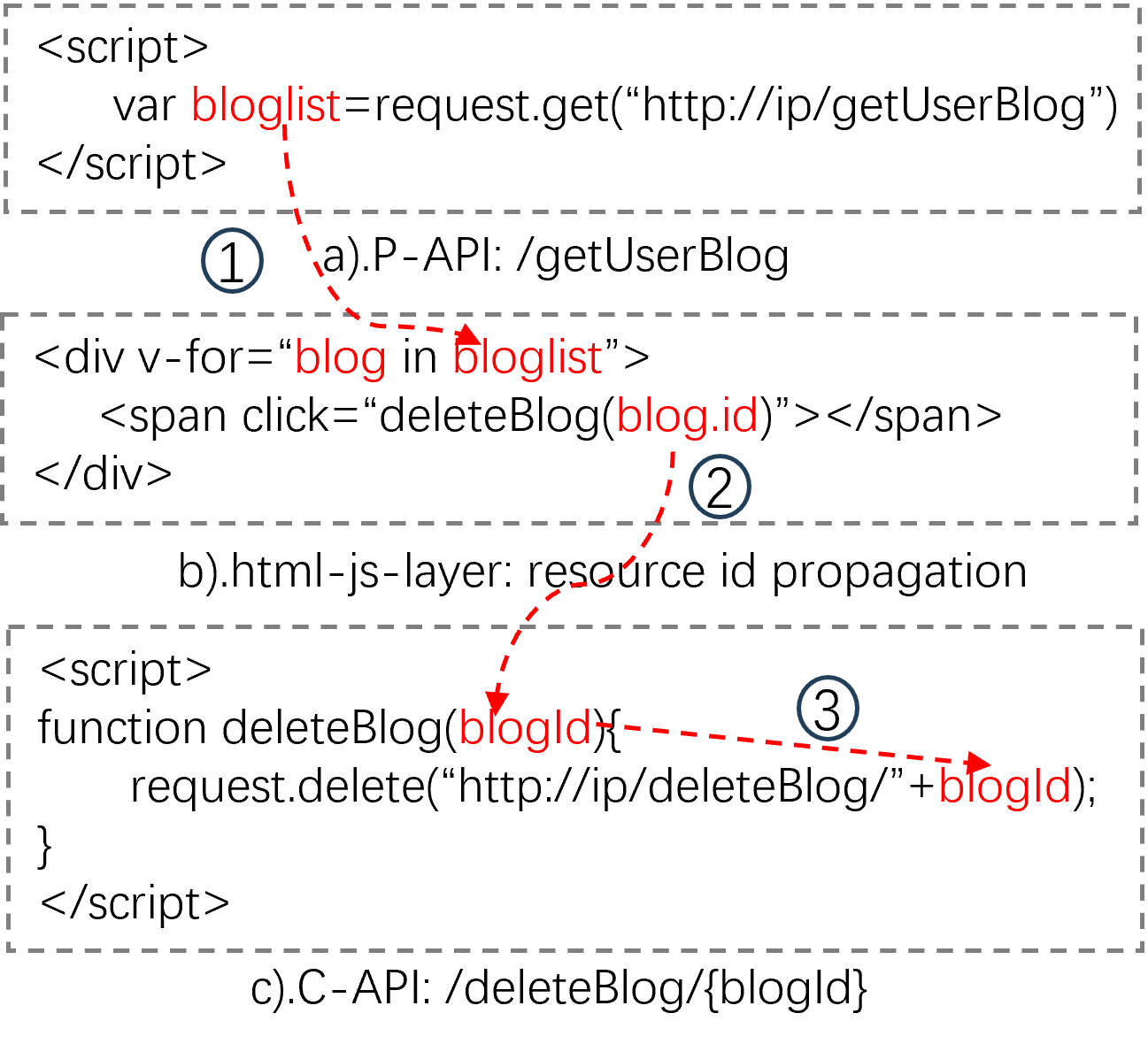}\label{fig_7_2}}
    \caption{Resource ID production, propagation and consumption}
    \end{figure}

    As shown in Fig.\ref{fig_4_1}, BOLAZ tracks the propagation of tokens in the API layer and SQL layer, and obtains the SELECT statements associated with the API return value. By analyzing the Abstract syntax tree (AST) of the SELECT statement, \tool determines that the SELECT statement obtains the primary key \emph{id} of the \emph{blog} table. Only when the associated SELECT statement obtains the resource ID columns of the table, \tool determines the API as P-API.

  \begin{figure}[htbp]
			\centering
			\includegraphics[scale=0.55]{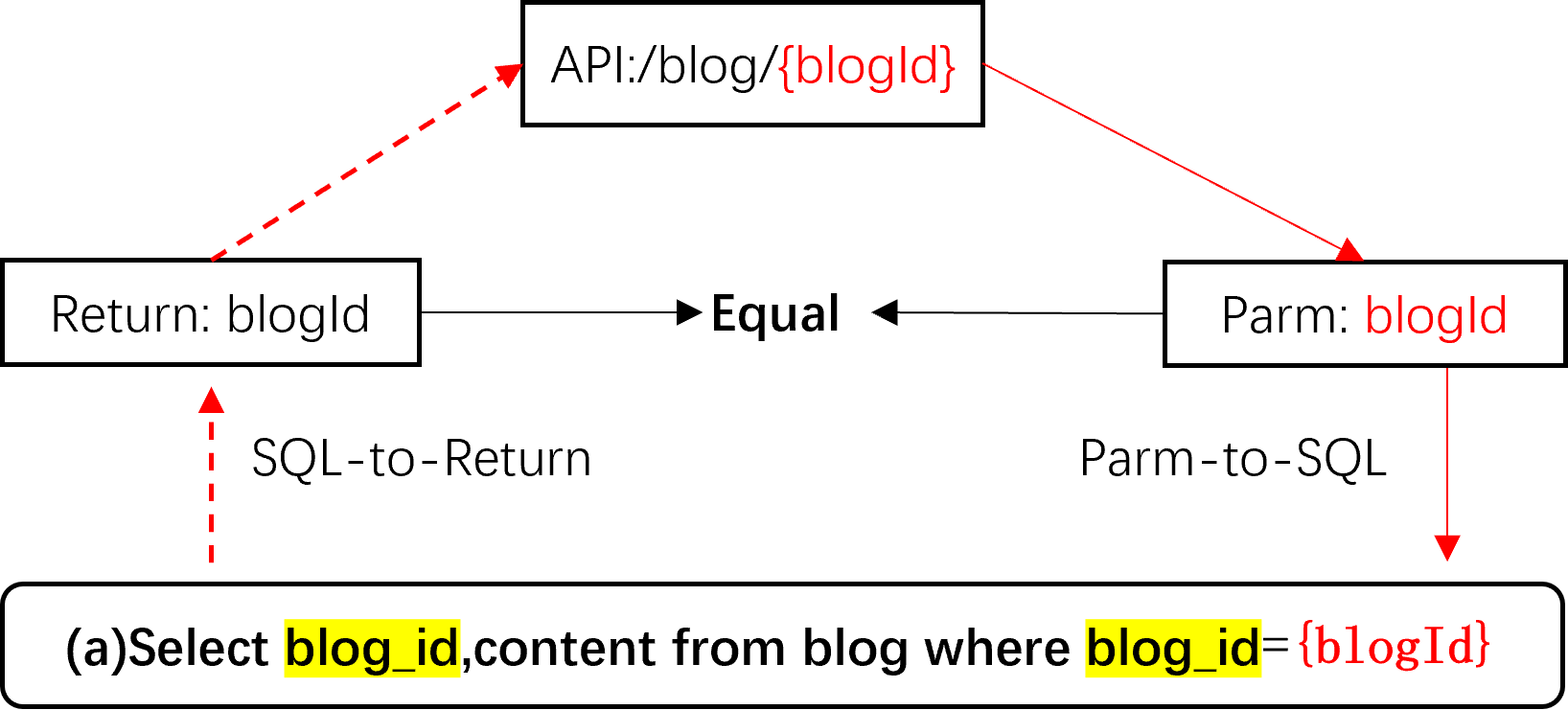}
			\caption{FP-API Identification. SQL-(a) produces the same \emph{blog ID} as the API parameter, not creating a new \emph{blog ID}. \tool considers the API for creating and consuming the same resource ID as a P-API, not an FP-API.}
			\label{fig_6_0}
	\end{figure}
 
	\subsubsection{MSG Interval Generation}
	Once the P-API is identified, the SELECT statements associated with the P-API return value are the MSG intervals. MSG intervals must maintain integrity to isolate resource IDs to the maximum authorization range. Additionally, since MSG intervals generated by P-APIs may have dependency conditions, \tool extracts these dependencies to ensure the accuracy of the authorization range.
    \begin{lstlisting}[
		caption=Reduction of MSG intervals,
		label=lst1]
Select a_id,addr from address where user_id="uid" and detail="userinput" limit start,stop
Select a_id,addr from address where user_id="uid"
	\end{lstlisting}
	
	\textbf{\emph{Integrity of the MSG interval.}}
	SELECT statements represent MSG intervals. Still, there may be some conditions in SELECT statements to reduce the range of MSG intervals. As shown in Listing \ref{lst1}(1), the SQL statement obtains the user's address. ``Detail" condition corresponds to the keyword search function of the address, and the user can make the detail condition cover the entire table without entering the keyword. ``Limit" corresponds to the address paging function, and users can also obtain all data through the page turning function. Conditions of this type will reduce MSG intervals. \emph{User\_id} is the foreign key of the \emph{addr} table, which is a necessary condition to determine MSG intervals. Therefore, \tool preserves only the foreign key and non-user input conditions in SELECT statements, as shown in Listing \ref{lst1}(2).
	
	\textbf{\emph{Dependence of the MSG interval.}} The complete MSG interval preserves the conditions of foreign keys, but these conditions are not necessarily independent. \tool obtains the dependency conditions of foreign keys from external inputs (API parameters and tokens) and internal associations (parent resource IDs) of API. 
	
\begin{figure}[htbp]
			\centering
			\includegraphics[scale=0.55]{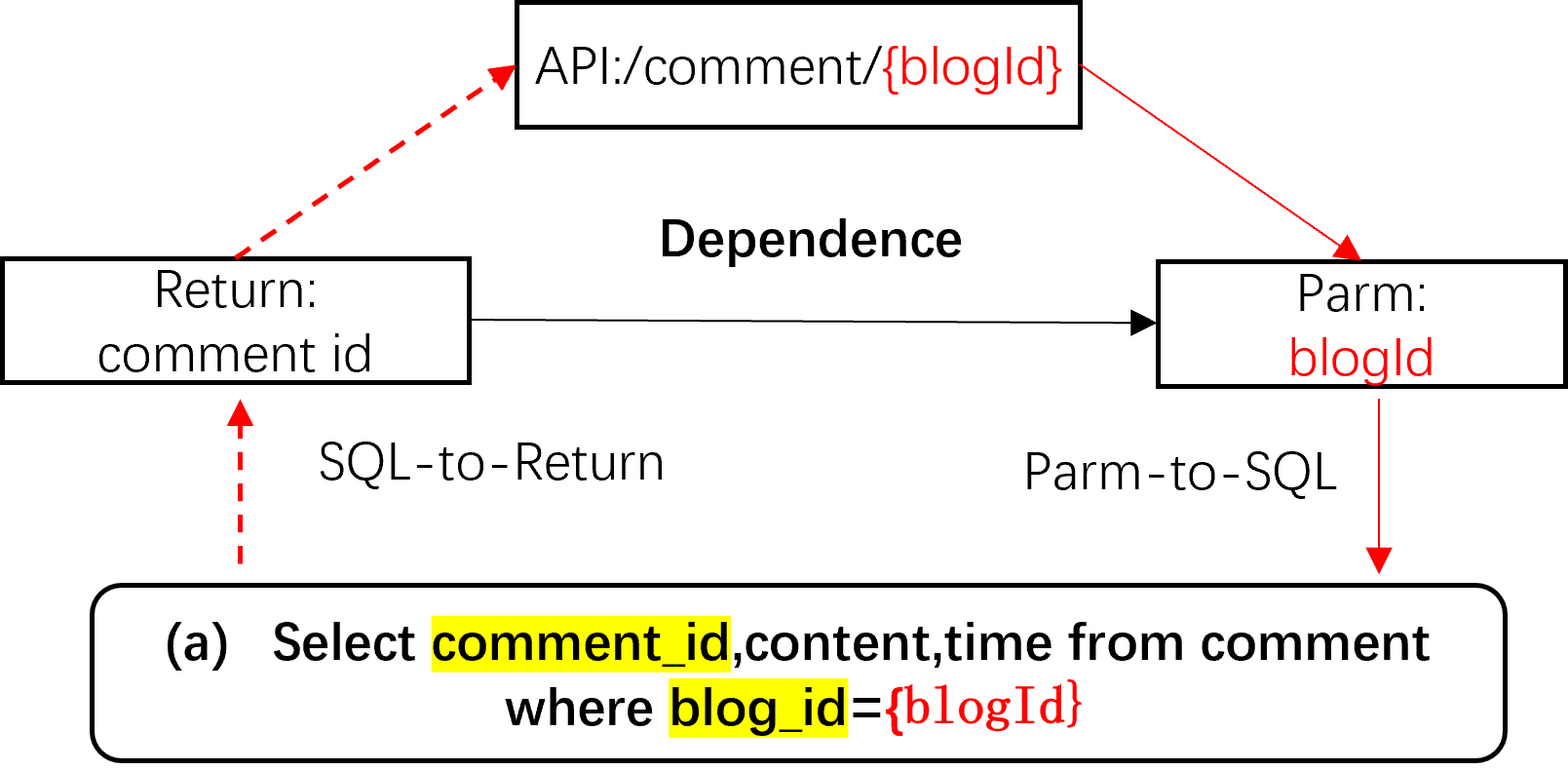}
			\caption{PC-API identification. the API generates \emph{comment IDs} while consuming the \emph{blog ID}, which is both a C-API and a P-API.}
			\label{fig_6}
	\end{figure}

	First, \tool analyzes the API parameters. The parameters of the P-API are propagated to the foreign key in the MSG interval, indicating that the P-API first consumes a specific type of resource ID and then produces another kind of resource ID. As shown in Fig.\ref{fig_6}, the \emph{comment ID}'s MSG interval depends on the \emph{blog ID}'s MSG interval. \tool tracks the propagation data flow of parameters to find the mapping relationship between parameters and foreign keys.
    
	Second, the API token represents the encrypted information of the user's own identity. Under the premise that the token is not leaked, the attacker is not capable of launching an attack by modifying the token. Therefore, tokens are not considered the potential injection points of BOLA attacks. However, after propagation, the token will be converted to the user ID, which may be passed to the conditions of SELECT statements in P-API. It is also a dependency condition for generating MSG intervals. \tool tracks the token propagation to discover whether tokens are associated with userId.
    \begin{figure}[htbp]
    \centering
			\includegraphics[scale=0.55]{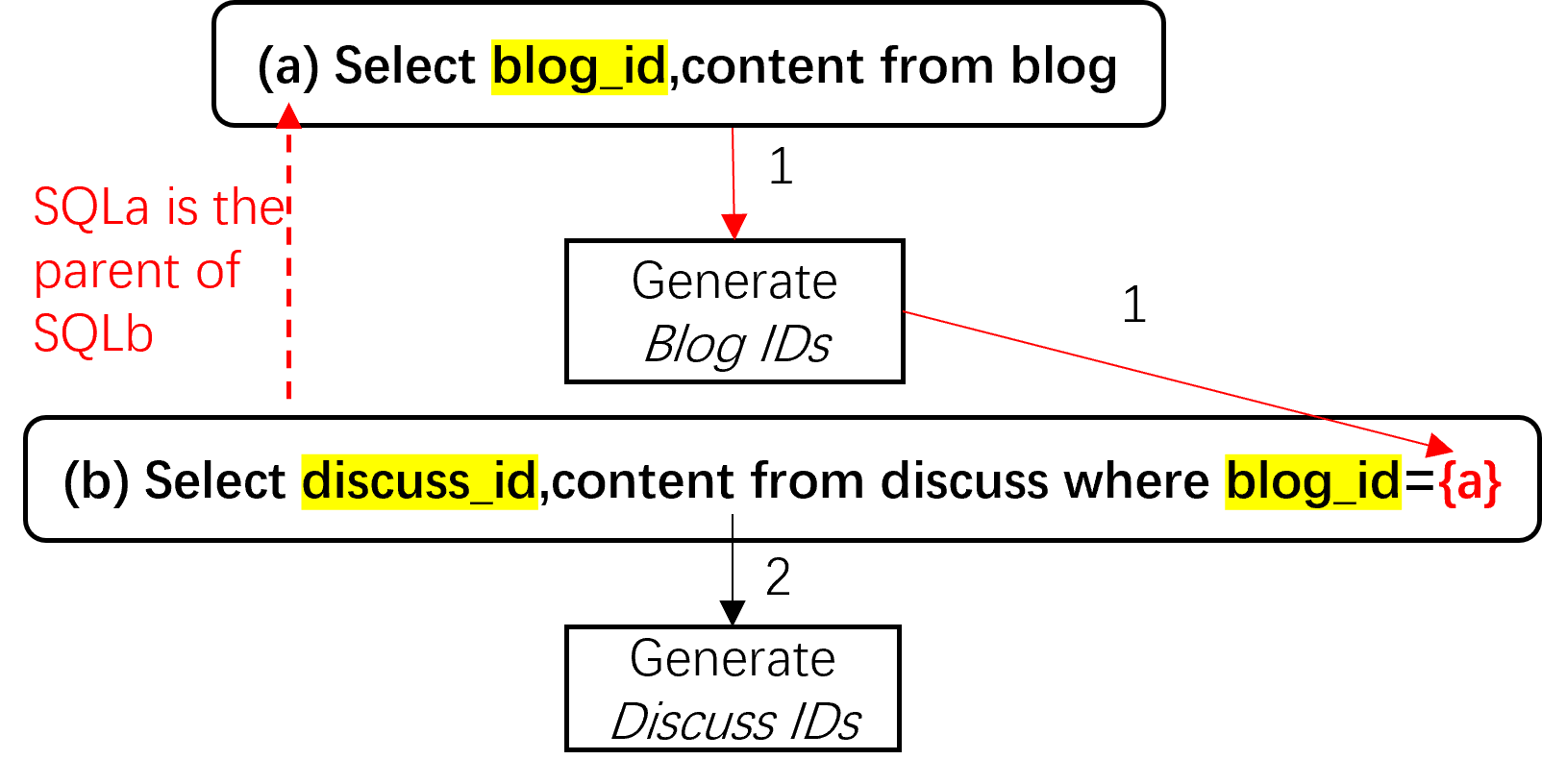}
			\caption{Dependency on resource ID}
			\label{fig_7}
    \end{figure}
	
	Third, there may be a parent-child dependency between resource IDs in P-APIs. Obtaining the MSG interval of the parent resource ID is the premise of calculating the MSG interval of the sub-resource ID. As shown in Fig.\ref{fig_7}, the API uses SQL-(a) and SQL-(b) to produce MSG intervals of \emph{blog ID} and \emph{discuss ID}, respectively. SQL-(a) for SQL-(b), and SQL-(b) depends on SQL-(a). \tool tracks the dependency of resource IDs within P-API between SQL statements.
	
	Finally, \tool identifies the P-API by analyzing the SELECT statement that returns the resource ID in the API and extracts MSG intervals.
	
	\subsection{C-API Identification and BOLA Attack Injection Point Discovery}
	C-API is downstream from the propagation data flow of resource ID. The BOLA attack takes the resource ID parameter of the C-API as the injection point, so \tool needs to identify the C-API and determine which parameters have a mapping relationship with the resource ID.

    BOLA attack injection points exist in user-controlled resource ID parameters, which eventually propagate to the SQL statement's primary or foreign key. Since the resource ID parameter does not change in the C-API propagation process (detailed in ~\ref{sec:Propagation invariance of resource IDs}), \tool also uses taint tracking technology \cite{kim_survey_2014} to analyze which API parameters are propagated to SQL statements, and determines which resource ID is associated with API parameters by analyzing the AST of SQL statements.
    
    As shown in Fig.\ref{fig_7_1}, \tool uses API parameters (\emph{blogId}) as taint sources and SQL statements as taint propagation endpoints. \tool found that \emph{blogId} eventually spread to the \emph{id} condition of Delete statement to delete \emph{blog}. \tool determines that the \emph{\/deleteBlog\/\{blogId\}} is C-API, and the \emph{blogId} parameter consumes the primary key \emph{id} of the \emph{blog} table.
    
    
    \tool determines whether the API parameters consume the resource ID from the context correlation between API parameters and primary or foreign keys in the four operation types of SQL statements, as shown in Table \ref{table1}.

        \begin{table}[htbp]
		\caption{C-API identification\label{table1}}
		\centering
		\begin{tabular}{lp{13cm}}
			\hline
			\textbf{Type} & \textbf{Mapping}\\
			\hline
			Select & There is a mapping relationship between API parameters and the where condition's primary key or foreign key in the SQL statement.\\
			\hline
			Delete & There is a mapping relationship between API parameters and the where condition's primary key or foreign key in the SQL statement.\\
			\hline
			Insert & API parameters are mapped with the primary or foreign key of the inserted value in the SQL statement.\\
			\hline
			$Update_{1}$ & There is a mapping relationship between API parameters and update values in SQL statements.\\
			\hline
			$Update_{2}$ & There is a mapping relationship between API parameters and the where condition's primary key or foreign key in the SQL statement.\\
			\hline
		\end{tabular}
	\end{table}

	\subsection{P-API and C-API Data Flow Association}
    Effective defense against BOLA attacks hinges on capturing API contexts \cite{teixeira_security_2023}.  In system development, developers retain the context relationship between P-APIs and C-APIs when writing front-end code. Therefore, \tool can also analyze the propagation relationship of resource IDs between P-APIs and C-APIs in front-end code through taint tracking technology.

    As shown in Fig.\ref{fig_7_2}. When the page is loaded, the Javascript code calls the P-API (\emph{/getUserBlog}) to get the \emph{blog} created by the user himself. The HTML code renders the \emph{blog} resources returned by the P-API and provides a \emph{click} event so that the user can call the \emph{deleteBlog} function. When the user initiates a click event in the page, the Javascript code calls the C-API (\emph{/deleteBlog/\{blogId\}}) to delete the corresponding \emph{blog} resource. \tool takes P-API as the starting point of taint tracking and C-API as the end point of taint tracking determines which P-API provides resource IDs for the C-API.
	
	\subsubsection{Identify The Data Flow of P-APIs And C-APIs}
\begin{figure}[htbp]
			\centering
			\includegraphics[scale=0.5]{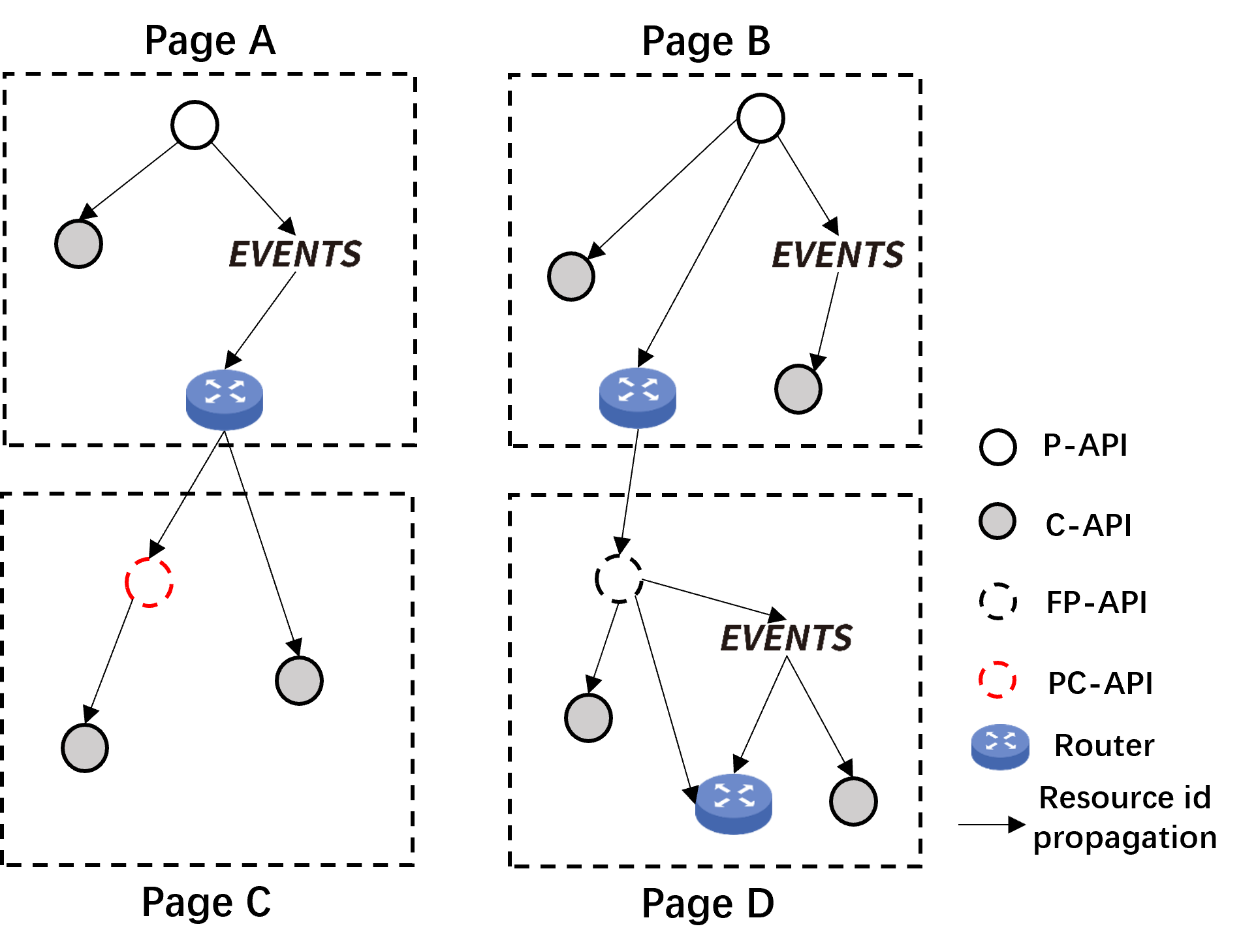}
			\caption{Data flow between P-API and C-API}
			\label{fig_8}
	\end{figure}

	\tool captures the propagation of resource ID within and between HTML pages. First, we analyze the propagation of resource ID within the HTML page. As shown in Fig.\ref{fig_8}, \tool divides the internal propagation of the page into four data flow patterns, with P-API as the starting point of data flow propagation. C-API will terminate the data flow after consuming the resource ID. The router is used to realize the function of jumping from one page to another, and the parameters can be carried out during the jumping process.
	\begin{itemize}
		\item{\emph{P-API to C-API}: P-API propagates resource IDs directly to C-API}
		\item{\emph{P-API to Event to C-API}: P-API propagates resource IDs to C-API through HTML events.}
		\item{\emph{P-API to Router}: P-API propagates resource IDs to Router.}
		\item{\emph{P-API to Event to Router}: P-API propagates resource IDs to Router through HTML events.}
	\end{itemize}
	
	Secondly, the propagation of resource ID between HTML pages can complement the data flow of P-API and C-API between multiple HTML pages. As shown in Fig.\ref{fig_8}, \tool uses two data flow patterns to receive the resource ID passed from the parent HTML page.
	
	\begin{itemize}
		\item{\emph{Router to PC-API}: Router receives resource IDs of parent pages and propagates them to PC-API.}
		\item{\emph{Router to C-API}: Router receives the resource ID of parent pages and propagates them to C-API.}
	\end{itemize}
	
	Through the above six data flow patterns, \tool tracking API workflows to obtain the logical relationship between attack injection points and MSG intervals, and ensuring that the authorization policy of injection points constantly changes with API workflows.
	
	\subsubsection{Generation of MSG Interval Set}
	The MSG interval of the resource ID produced by the P-API is the authorization interval of the associated C-API attack injection point. However, the three characteristics of MSG intervals generated by P-APIs lead to the limitations of the authorization interval of attack injection points in C-APIs: the diversity of resource IDs, the dependence between resource IDs, and the incompleteness of foreign key resource IDs. Therefore, \tool performs the following three operations on MSG intervals.
	
	\textbf{\emph{Matching.}} P-API may create  MSG intervals for multiple resource IDs. Therefore, \tool must match the resource ID consumed by C-API with the resource ID created by P-API.
	
	\textbf{\emph{Dependence Resolving}.} When there is any dependency between resource IDs, \tool will analyze the MSG interval of the parent resource ID. If the MSG interval of the parent resource ID covers all the data in the database table, the child resource ID will discard the limitation of the parent resource ID. Otherwise, \tool will update the condition of the parent resource ID to the MSG interval of the parent resource ID.
	
	\textbf{\emph{Backtracking.}} The primary or foreign key obtained by P-API may become the resource ID that C-API requires. The MSG interval of the primary key is complete, which can provide the C-API with a full range of permissions for the available resource ID. The foreign key only has a partial mapping relationship with the primary key of other resources, and the generated MSG interval is incomplete.
	
\begin{figure}[htbp]
    \centering
			\includegraphics[scale=0.55]{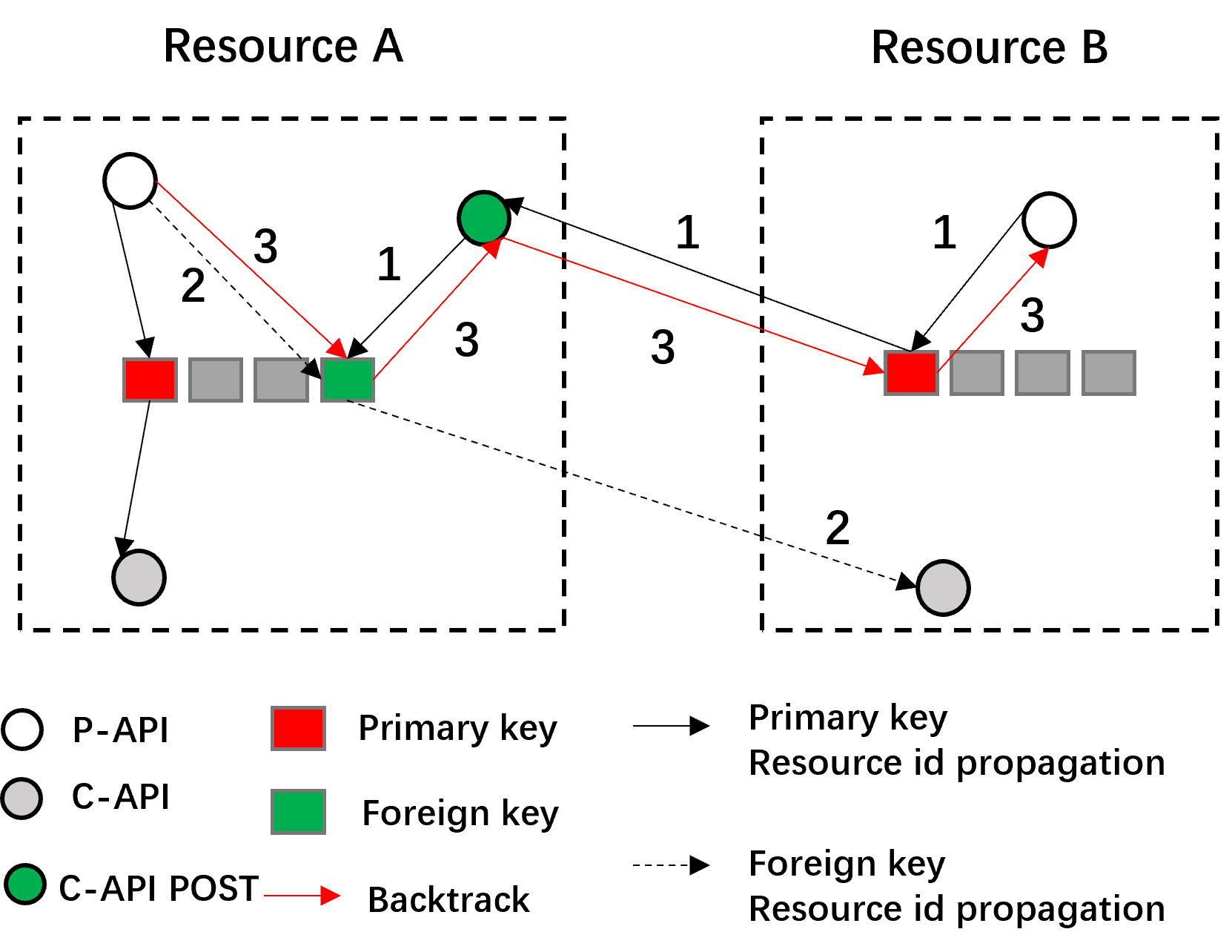}
			\caption{Data flow backtracking}
			\label{fig_9}
    \end{figure}

	The foreign key is inserted into the database by the C-API of the resource creation type (POST), so the MSG range of the foreign key is created by the P-API that provides the resource ID for the POST C-API. As shown in Fig.\ref{fig_9}, in data stream 1, the P-API of Resource B passes the resource ID of the primary key to the POST C-API of Resource A, and the C-API stores the primary key of Resource B as the foreign key of Resource A. In data stream 2, the P-API propagates the resource ID of the foreign key to the C-API. The MSG interval of the foreign key resource ID in data stream 2 comes from the P-API in data stream 1. Therefore, \tool converts the MSG interval of the foreign key into the MSG interval of the primary key of the P-API associated with the POST C-API through data stream backtracking, as shown in data stream 3.
	
	C-API may have multiple associated P-APIs, which means it is related to numerous MSG intervals, thus forming a set of MSG intervals.
	 
	\subsubsection{Optimization of MSG Interval Set}
	There may be subset or intersection relationships between the internal elements of the MSG interval set, and performing repeated interval queries will result in considerable performance costs. Therefore, \tool defines the merging rules of SELECT statements to reduce its number under the premise of ensuring that the MSG interval does not change, thus reducing the number of SELECT statements and the query of repeated intervals.
	
\begin{figure}[htbp]
			\centering
    		\includegraphics[scale=0.5]{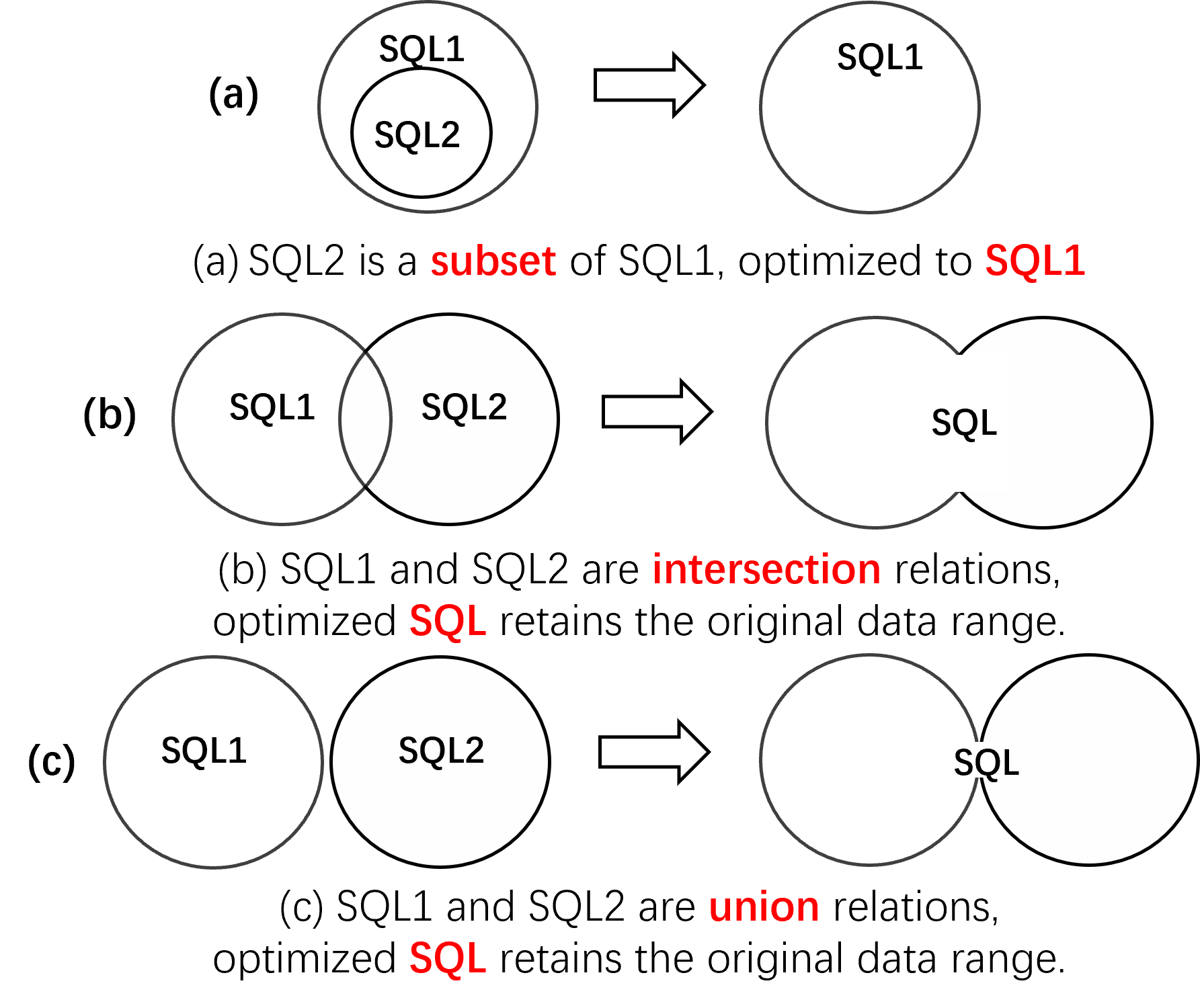}
    		\caption{MSG interval optimization}
    		\label{fig_10}
	\end{figure}

	First, we define the identifiers in the merging rules, as shown in Table \ref{table2}. Then, we discuss the merging rules of SQL statements from the perspective of the coincidence relationship between MSG intervals, As shown in Fig.\ref{fig_10}.
    	\begin{table}[htbp]
		\caption{Merge rule identifier\label{table2}}
		\centering
		\begin{tabular}{p{1cm}p{9cm}}
			\hline
			\textbf{Type} & \textbf{Meaning}\\
			\hline
			$S_q$ & Single table query.\\
			\hline
			$M_q$ & Multi table query.\\
			\hline
			$M_t$ & The main table in multi-table query.\\
			\hline
			$S_i$ & The ith MSG interval (SQL statement) associated with the C-API.\\
			\hline
			$T_i$ & $T_i$ is the table corresponding to resource ID produced in $S_i$.\\
			\hline
			$W_i$ & $W_i$ is the set of where conditions related to $T_i$ in $S_i$.\\
			\hline
		\end{tabular}
	\end{table}
	
	\textbf{\emph{Rule\#1:Subset rule.}} When there is a subset relationship between the two SQL statements, the \tool merges the two SQL statements into a parent SQL statement.
	
	\emph{Precondition:} 
	\begin{itemize}
		\item $ S_i \in S_q $ , $ S_j \in S_q $ , $ T_i = T_j $ 
		\item $ S_i \in M_q $ , $ S_j \in M_q $ , $ T_i \in M_t $ , $ T_j \in M_t $ , $ T_i = T_j $
		\item $ S_i \in S_q $ , $ S_j \in M_q $ , $ T_j \in M_t $ , $ T_i = T_j $ 
	\end{itemize}
	
	\emph{if:}
    
    \qquad$ W_i = \emptyset $  or $ W_i \subseteq W_j $  
    
	\emph{then:} 
    
    \qquad$ S_i \cup S_j = S_i $
	
	\textbf{\emph{Rule\#2:Intersection and union rule.}} When two SQL statements have an intersection or union relationship, the \tool merges the where conditions of the two SQL statements.
	
	\emph{Precondition:}  $ S_i \in S_q $ , $ S_j \in S_q $ , $ T_i = T_j $
	
	\emph{if:} 
    
    \qquad $ W_i \cap W_j = \emptyset $ or $ W_i \cap W_j \ne \emptyset $ , $ W_i \not\subseteq W_j $ 
	
	\emph{then:} 
    
    \qquad $ S_i \cup S_j = W_i \cup W_j $
 
	\tool uses the optimized MSG interval to perform a resource ID authorization check on the BOLA attack injection point of the C-API. Suppose the optimized MSG interval can cover all the database table data. In that case, it shows that the resource ID consumed in the C-API is unlimited, and the \tool does not perform authorization checks on the C-API.

\section{Implementation}\label{implementation}

\begin{figure}[!htbp]
        \centering
    		\includegraphics[scale=0.5]{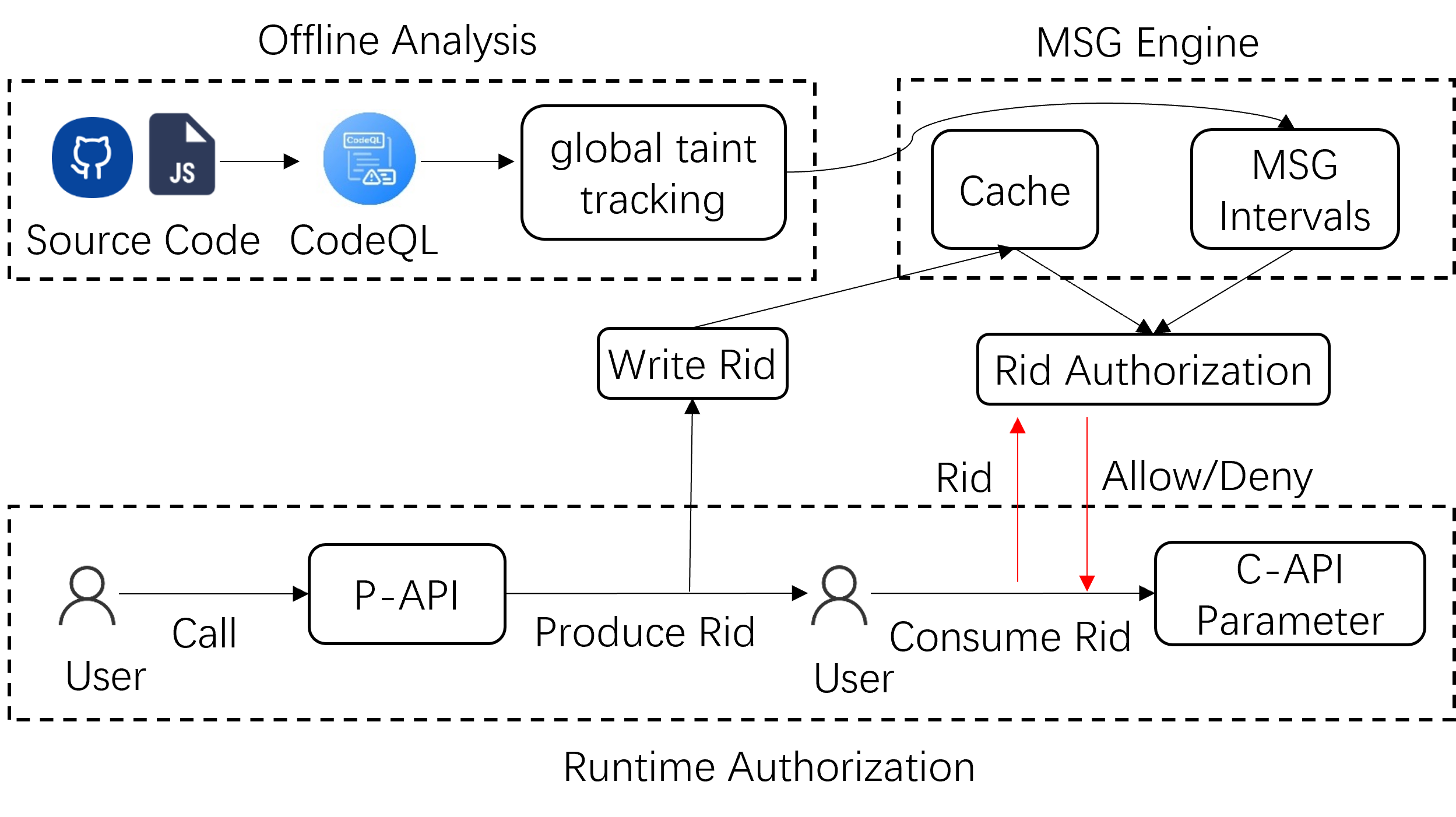}
    		\caption{Implementation of the \tool}
    		\label{fig_11}
    \end{figure}

In this section, we detail the implementation of \tool, a tool for detecting BOLA vulnerabilities in SpringBoot-based web applications. As shown in Fig.\ref{fig_11}, \tool includes offline analysis and runtime authorization.

	\subsection{Offline Analysis}
    \tool utilizes CodeQL's taint tracking tool \cite{codeql_codeql_2024} to trace the data flow of resource IDs. \tool focuses on determining MSG intervals for APIs that operate under ordinary user permissions. If an attacker gains administrator privileges, he has permission to operate on all resources and does not need to launch BOLA attacks. Not performing authorization checks on administrator rights APIs also reduces the performance overhead of \tool.
    
    The first step is identifying P-APIs. \tool uses global taint tracking \cite{codeql_global_2024}, treating API parameters or function expressions as the source and API returns as the sink. During the data propagation process, \tool analyzes the abstract syntax tree (AST) of SQL statements to extract resource IDs and determine the MSG intervals produced by the API. The generated MSG intervals are stored in the MSG engine. \tool also tracks token propagation. While RESTful APIs typically use token-based authentication methods such as JWT \cite{jwt_json_2015} or OAuth \cite{dick_hardt_oauth_2012}, the specific implementation varies across applications, making it challenging to extract token flags from source code. Therefore, \tool requires developers to provide the relevant token code flags.

    The second step is C-API identification. \tool uses API parameters as the source and SQL statements as the sink to map the relationship between parameters and resource IDs, identifying both BOLA attack injection points and C-APIs.

    Finally, to establish the data flow associations between P-APIs and C-APIs, \tool analyzes the propagation paths of resource IDs in the front-end code. By leveraging six data flow patterns, it identifies direct or indirect resource ID propagation throughout the system.
 
	\subsection{Runtime Authorization}
	\tool checks the authorization of resource ID according to MSG intervals generated by offline analysis when the system is running. Resource IDs used by users must exist in MSG intervals, otherwise, the C-API call will be blocked. We consider that when the database data size is large, if database queries are performed at each C-API authorization check, it will lead to high latency. C-APIs will be called in a short time after users call P-APIs. Hence, the MSG engine caches resource IDs produced by P-APIs.  If the cache is not hit, a database query is performed. A cache or database hit means the C-API has access to the resource ID.

\section{Evaluation}\label{evaluation}

\subsection{Research Questions}
	\noindent\textbf{\emph{RQ1: How is the effectiveness of \tool identifying API types and associations?}}
	
	\noindent First, \tool transforms the defense of BOLA attacks into three sub-problems in the resource ID workflow(see ~\ref{sec:ProblemDefinition}). Hence, our experiments first evaluate the effectiveness of \tool by verifying the technical consistency of each `module' (see ~\ref{sec:Overview}) through data flow analysis.
     
	\noindent\textbf{\emph{RQ2: How much extra performance overhead does \tool cost?}}
	
	\noindent Second, \tool performs authorization checking on resource IDs by caching or database queries, which results in extra performance overhead. We evaluate the performance overhead of \tool by comparing the original system and the \tool-added system in three aspects: P-API storage cache, C-API cache query and database query.
	
	\noindent\textbf{\emph{RQ3: How effective is \tool's defense against real-world BOLA attacks?}}
	
	\noindent \tool should be able to defend against BOLA attacks in the real world. We first use the CVE vulnerability to evaluate the effectiveness of \tool's defense against existing vulnerabilities. Second, we verify \tool's ability to detect unknown vulnerabilities by scanning the BOLA vulnerability of open-source projects.

    \noindent\textbf{\emph{RQ4: How does \tool compare with the SOTA approach?}}
    
    \noindent BOLARAY is the most closely relevant work to \tool and is the SOTA method for detecting BOLA vulnerabilities. Therefore, we compared the vulnerability detection effect of BOLARAY and \tool in real-world projects.
 
	\subsection{Dataset}
         \begin{table*}[htbp]
		\caption{Statistics on evaluation projects\label{table3}}
		\centering
		\begin{tabular}{p{6cm}rrrrl}
			\hline
			\textbf{Project} & \textbf{Stars} & \textbf{Files} & \textbf{LLOC} & \textbf{APIs} & \textbf{Description}\\ 
			\hline
			Blog-master \cite{mqpearth_blog_2024} & 616 & 106 & 42,366 & 57 & Blog System\\
			\hline
			BookStore-master \cite{stuhaibin_bookstore_2020} & 314 & 168 & 29,944& 71 & Bookstore Project\\
			\hline
			Mall-master \cite{macrozheng_mall_2024} & 78.8k & 121 & 15,467 & 65 & E-commerce System\\
			\hline
            NewbellMall-master \cite{newbee-ltd_newbee-mall_2024} & 1.4k & 140 & 16,546 & 62 & Mall System\\
            \hline
            IceCms-master \cite{thecosy_icecms_2024} & 1.7k & 241 & 80,308 & 109 & Content Management System\\
            \hline
            MusicwWbsite-master \cite{yin-hongwei_music-website_2024} & 5.7k & 133 & 42,906 & 63 & Music Website\\
            \hline
            OnlineExam-master \cite{yxj2018_springboot-vue-onlineexam_2024} & 2k & 106 & 8,116 & 46 & Online Examination System\\
            \hline
            UniversityForum-master \cite{cp3geek_universityforum_2022} & 196 & 64 & 13,279 & 16 & University Campus Forum\\
            \hline
            InformationSystem-master \cite{boylegu_informationsystem_2023} & 2k & 20 & 2,344 & 4 & People Information System\\
            \hline
            OnlineMall-master \cite{tablu666_onlinemall_2020} & 32 & 118 & 10,640 & 33 & Online Mall\\
            \hline
			Total & - & 1,217 &261,916 & 526 & -\\
			\hline
		\end{tabular}
	\end{table*}

    \subsubsection{How to collect projects}
    As the implementation of \tool is on the SpringBoot framework in Java language, we searched for repositories that depend on SpringBoot on GitHub \cite{spring-projects_spring-boot_2025}, SourceCodeExamples \cite{sourcecodeexamples_sourcecodeexamples_2025} and LibHunt \cite{libhunt_open-source_2025} and filtered them by application types, authentication modes, and logic control modes.
    
    
    The selected 10 projects cover various application types, two authentication modes, and two logic control modes. 1). Application type. There are similarities in the business types of open-source projects. The project we selected contains Blog, Bookstore, E-commerce, Content Management System, Online Examination System, People Information System, etc. We screen systems with different business logic so that \tool can cover a variety of business scenarios. 2). User authentication mode. User authentication modes include tokens and cookies, so the system we have chosen covers both user authentication modes. 3). Logical control modes (detailed in ~\ref{sec:MSG Rules Of Resource ID}), as shown in Table \ref{table3}. Based on these three, the selected projects are sufficient and can effectively cover our scope of problems. We focus on the API data flow, which requires a good coverage of business types. It is of little significance to involve more projects. 

	\subsubsection{RQ1 \& RQ2} There is no tool to achieve the classification and association of API. To construct the dataset's ground truth, we need to deploy projects locally and manually count and identify P-API, C-API and the relationship. The whole process is complex and time-consuming. So we selected five systems from Table \ref{table3} based on application types, logical control mode, and user authentication modes to contribute a benchmark to verify the effectiveness of \tool, as shown in Table \ref{table4}. Sufficient to verify the effectiveness of \tool.
 
	We manually analyzed the API categories and relationships to build the ground truth data. First, we deployed and ran the project locally, logging in as a regular user to access all functionalities. Simultaneously, we used Fiddler \cite{telerik_fiddler_2024} to capture the API list. Next, we examined the API server’s source code. Given that the current architecture follows a three-tier model, we were able to quickly categorize the APIs. We then analyzed the front-end source code to trace the propagation paths of resource IDs both within and across pages. Our team was divided into two groups, analyzing the project’s source code independently, followed by cross-validation of the results. Across the five projects, we identified a total of 40 P-APIs and 54 C-APIs.

    \subsubsection{RQ3-1} We searched the CVE vulnerability database for BOLA (a.k.a., IDOR) vulnerabilities. The results indicate that only a few are from Java projects (e.g., SpringBoot). Finally, we screened out three vulnerabilities corresponding to the three modes of BOLA attack from the CVE-2023-36100 \cite{cve_cve-2023-36100_2023} and CVE-2023-32310 \cite{cve_cve-2023-32310_2023}.
    
    \subsubsection{RQ3-2} 
    The survey of existing work \cite{huang_detecting_2024}, \cite{bocic_finding_2016}, \cite{son_rolecast_2011}, \cite{monshizadeh_mace_2014} found no current large-scale benchmarks. We detect unpublished BOLA vulnerabilities in these 10 projects to evaluate the effectiveness of \tool in the real world.

    \subsubsection{RQ4}
    BOLARAY is used to detect BOLA vulnerabilities in PHP. We adapted BOLARAY to the SpringBoot framework and compared it with \tool by detecting BOLA vulnerabilities in the collected 10 projects.
    \begin{table}[htbp]
    \setlength\tabcolsep{4.5pt}
		\caption{Benchmark of Identifying API Types and Associations\label{table4}, R denote Relation, LCM denote Logic control mode, UAM denote User authentication mode.}
		\centering
		\begin{tabular}{p{3cm}rrrll}
			\hline
			\textbf{Project} & \textbf{P-API} & \textbf{C-API} & \textbf{API-R} & \textbf{LCM} & \textbf{UAM}\\ 
			\hline
			NewbellMall & 9 & 14 & 17 & Server & Token\\
			\hline
			OnlineExam & 11 & 20 & 20& Server & UserId\\
			\hline
			Blog & 10 & 14 & 49 & Client & Token\\
			\hline
            UniversityForum & 9 & 4 & 10 & Server & UserId\\
            \hline
            InformationSystem & 1 & 2 & 2 & Server & Token\\
            \hline
			Total & 40 & 54 &98 & - & -\\
			\hline
		\end{tabular}
	\end{table}
	
	\subsection{Metrics}
	In our experiment, we evaluated the following indicators. We use Precision and Recall to measure the effectiveness of API classification and API association.
	True Positive(TP) represents the number of correctly classified APIs; The number of API pairs that are correctly associated.
	False Positive (FP) represents the number of APIs classified as P-API (C-API) but not P-API (C-API); The number of API pairs that are judged to be associated with P-APIs and C-APIs but do not exist. False Negative (FN) represents the number of APIs classified as not P-API (C-API) but P-API (C-API); The number of P-API and C-API pairs that are judged not to be associated but are associated. The formulas for  Precision (Pr) and Recall (Re) are as follows.
	\begin{equation}
		Pr = \frac{TP}{TP+FP}, Re = \frac{TP}{TP+FN}
	\end{equation}

	\subsection{RQ1: How is the Effectiveness of \tool Identifying API Types and Associations?}
 
    \subsubsection{Setup} The experiment was completed in two steps. First, we use \tool to classify the APIs in the project. Suppose there is an error in the API classification. In that case, we will modify the API to the correct type because the premise of accurately associating the API is to obtain the accurate API type. In the second step, we use \tool to associate P-API with C-API. 

	\begin{table}[htbp]
        \setlength\tabcolsep{5.5pt}
		\caption{Effectiveness of API Classification and Association\label{table5}, P denotes P-API, C denotes C-API, R denotes Relation.}
		\centering
		\begin{tabular}{p{3cm}cccccc}
			\hline
			\textbf{Project} & \textbf{P-Pr} & \textbf{P-Re} & \textbf{C-Pr} & \textbf{C-Re} & \textbf{R-Pr} & \textbf{R-Re}\\
			\hline
			NewbeeMall & 8/8 & \textcolor{red}{8/9} & 14/14 & 14/14 & 17/17 & 17/17\\
			\hline
			OnlineExam & 11/11 & 11/11 & 20/20 & 20/20 & \textcolor{red}{20/21} & \textcolor{red}{18/20}\\
			\hline
			Blog & 10/10 & 10/10 & 13/13 & \textcolor{red}{13/14} & 42/42 & \textcolor{red}{42/49}\\
			\hline
            UniversityForum & 9/9 & 9/9 & 4/4 & 4/4 & 7/7 & \textcolor{red}{7/10}\\
			\hline
            InformationSystem & 1/1 & 1/1 & 2/2 & 2/2 & 2/2 & 2/2\\
			\hline
		\end{tabular}
	\end{table}
	
	\subsubsection{Results} The experimental results of \tool are reported in Table \ref{table5}. 
    BolaZ achieved recall rates of 97\% for API classification and 87\% for API association, with one false positive. The precision of API classification and association is high because the resource ID, as the unique identification of the resource, is rarely modified in the process of data flow propagation. \tool can categorize and associate APIs very accurately without interrupting the taint tracking. We also analyze the reasons for the failure of API classification. By examining the log, we found that CodeQL generated data flow interruption at the \emph{Arrays.asList} method during the taint tracking process, failing P-API identification of the Newbellmall project. The Blog project also failed C-API recognition due to data flow interruption. 
	
	We observed that the Online-Exam, UniversityForum and Blog projects have low API association recall rates. First, we analyze Online-Exam and UniversityForum projects. The C-APIs of association failure are all attack injection points of userId, such as \emph{GET /api/score/\{current\}/\{size\}/\{studentId\}}. By analyzing the source code, we found that no P-API provides studentId for C-APIs that fail to correlate, and these studentId are derived from the client's local cookie. The Online-Exam Project saves the \emph{student ID} to the client's cookie after the student user logs in and provides it to the C-API. REST APIs typically use encryption, such as JWT, for authentication, and the Online Exam project's method of using the \emph{user ID} is not secure. At present, the \emph{user ID} will exist in more API return data, and it is easy for attackers to use the \emph{user ID} to launch BOLA attacks. Although \tool can solve this problem by extending the data flow from client data storage (cookie, local storage) to C-API, this API authentication mode is not secure.

    \tool generated the only false positive in analyzing API relationships in the online-exam project. After analyzing the source code, we found that the reason was that the client code modified the resource ID generated by P-API and passed it to the downstream C-API, resulting in \tool incorrectly associating the context of P-API and C-API. BOLAZ needs to analyze the source code to obtain the modification logic of the resource ID to solve this type of false positive. The problem is transformed into extracting operational semantics from the source code.
    
	Secondly, we analyze that the Blog project is due to the existence of client logic control in the system, which leads to the failure of the API association. For example, \emph{v-if= getStoreName() == name \textbar\textbar getStoreRoles().indexOf("ADMIN") \textgreater -1}. This condition means that the user's name stored by the client is equal to the user name of the resource, or the stored permission string contains ADMIN so that the user can see the button to delete the blog and have the right to call the API to delete the blog. However, \tool cannot accurately determine the meaning of this condition, failing the association between the deletion API of blog resources and the list API.
    
        \begin{tcolorbox}[colframe=black!50,colback=gray!10,left=1mm,right=1mm,top=0.5mm,bottom=0.5mm]
    {\textbf{Insight \#1:} Static analysis combined with dynamic analysis is a worth trying direction, which has the potential to refine \textbf{propagation logic} of resource ID.}
    \end{tcolorbox}

      \begin{tcolorbox}[colframe=black!50,colback=gray!10,left=1mm,right=1mm,top=0.5mm,bottom=0.5mm]
    {\textbf{Answer to RQ1:} \tool achieved recall rates of 97\% for API classification and 87\% for API association, with few false positives.}
    \end{tcolorbox}
	
	\subsection{RQ2: How Much Extra Performance Overhead does \tool Cost?}
	
	\subsubsection{Setup} To evaluate the performance overhead caused by \tool, we select the Newbeemall project as the test system of performance overhead from the GitHub stars, the number of APIs, and the recall of identifying and associating APIs, and select P-APIs and C-APIs, as shown in Table \ref{table6}. 
 
    We conducted a round-trip latency (RTT) comparison test between the original application system and the \tool-added system to evaluate the extra performance overhead of \tool, which mainly includes three aspects: the extra RTT brought by P-APIs cache resource ID; the extra RTT brought by C-APIs query cache; the extra RTT brought by C-APIs query database. 

	\begin{table}[htbp]
		\caption{Test API\label{table6}}
		\centering
		\begin{tabular}{lp{4.27cm}l}
			\hline
			\textbf{Type} & \textbf{API} & \textbf{Associated P-API}\\
			\hline
			P-API & GET /address & -\\
			\hline
			P-API & GET /shop-cart/page & -\\
			\hline
			P-API & GET /order & -\\
			\hline
			C-API & GET /order/\{orderNo\} & GET /order\\
			\hline
			C-API & DELETE /shop-cart/\{cartItemId\} & GET /shop-cart/page\\
			\hline
			C-API & DELETE /address/\{addressId\} & GET /address\\
			\hline
		\end{tabular}
	\end{table}

	\subsubsection{Results} 
\begin{figure}[htbp]
			\centering
    		\includegraphics[scale=0.4]{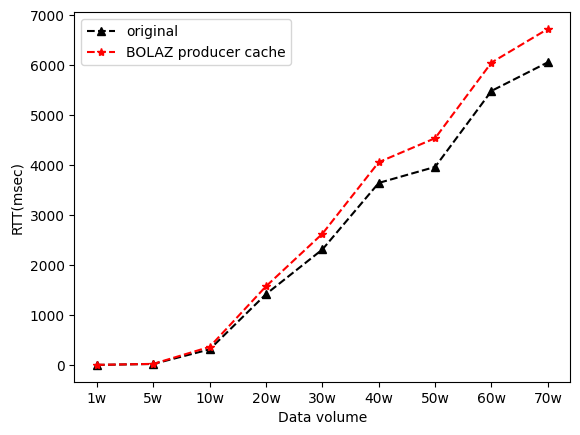}
            \captionsetup{font={small}}
    		\caption{Performance overhead of P-API}
    		\label{fig_12_1}
    \end{figure}

Firstly, we test the performance cost of the P-API cache resource ID. To restore the real application scenario, we use Apache JMeter \cite{jmeter_jmeter_2024} to simulate 100 users accessing multiple P-APIs concurrently at different data levels and average the RTT. The test results are shown in Fig.\ref{fig_12_1}. As the amount of data in the database increases, the performance cost of \tool increases. When the original system has 6 seconds RTT, \tool only increases the delay by hundreds of milliseconds, which users can accept.
	
\begin{figure}[htbp]
			\centering
    		\includegraphics[scale=0.4]{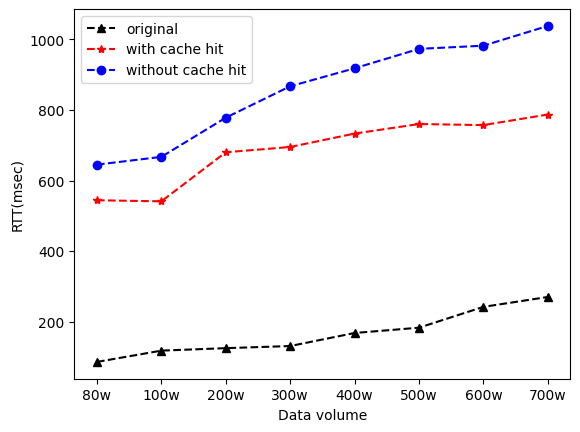}
            \captionsetup{font={small}}
    		\caption{Performance overhead of \tool}
    		\label{fig_12_2}
	\end{figure}

	Second, we test the performance overhead caused by \tool's MSG interval checking of C-API. We first tested the average RTT of C-API at different data levels under 1,000 user concurrency in the original system. Then, we tested the average RTT of system cache hits and misses under \tool protection. The test results are shown in Fig.\ref{fig_12_2}. Through experiments, we found that 1) The delay caused by \tool to the original system remains within 1 second whether it is a cache hit or not. 2) The cache hit query speed is faster than the direct query database, but the gap is not very big. The gap is small because \tool adds the primary key resource ID as a query condition to the SQL statement. After testing in Mysql8, the query speed of SQL statements with primary key conditions can be maintained at the millisecond level at the 700w data level. 
    
    The performance overhead of BOLAZ is due to the query cache or database, independent of original systems. The overhead is entirely related to the speed of cache or database queries. 
    \begin{tcolorbox}[colframe=black!50,colback=gray!10,left=1mm,right=1mm,top=0.5mm,bottom=0.5mm]
    {\textbf{Answer to RQ2:} Regardless of P-APIs cache resource IDs or C-APIs query resource IDs from the cache and database, the performance cost of \tool remains at the millisecond level.}
    \end{tcolorbox}
	
	\subsection{RQ3:How Effective is \tool's Defense Against Real-world BOLA Attacks?} 
     \begin{table}[htbp]
		\small
		\caption{Evaluation on BOLA vulnerabilities\label{table7}}
		\centering
		\begin{tabular}{|l|c|l|}
			\hline
			\textbf{Type} & \textbf{API Type} &\textbf{API} \\
			\hline
			\makecell[l]{\multirow{2}*{BFLA}} & \makecell[c]{Vertical Privilege} & GET /squareComment/getAllSquare/\{page\}/\{limit\} \\
			\cline{2-3} & \makecell[c]{CVE-2023-36100(C-API)} & POST /api/User/ChangeUser/\{jwt\} Body:\textcolor{red}{userId} \\
			\cline{2-3} & \makecell[c]{P-API} &GET /User/GetUserInfoByid/\{user\_self\_id\} \\
			\hline
			\makecell[l]{\multirow{2}*{BOPLA}} & \makecell[c]{Excessive Data Exposure} & GET /api/share/treeList \\
			\cline{2-3} & \makecell[c]{CVE-2023-32310(C-API)} & POST /api/share /removePanelShares/\{\textcolor{red}{panelId}\} \\
			\cline{2-3} & \makecell[c]{P-API} & \makecell[l]{POST /api/share/shareOut} \\
			\hline
			\makecell[l]{\multirow{2}*{UASBF}} & \makecell[c]{CVE-2023-32310(C-API)} & POST /api/sys\_msg/batchDelete Body:\textcolor{red}{msgId} \\
			\cline{2-3} & \makecell[c]{P-API} & \makecell[l]{POST /api/sys\_msg/list/\{goPage\}/\{pageSize\}} \\
			\hline
		\end{tabular}
	\end{table}
    \subsubsection{Results}
    The results of the evaluation on real-world BOLA vulnerabilities are shown in Table \ref{table7}.
    
    \textbf{BFLA(CVE-2023-36100).} The attacker obtains the userids of all users by accessing API1\emph{(GET /squareComment/getAllSquare)} vertically unauthorizedly and then tampering with the \emph{userId} injection point of C-API1\emph{(POST /api/User/ChangeUser Body:User,userId)} to modify the information of other users. By analyzing the data flow, \tool found that P-API1\emph{(GET /User/GetUserInfoByid/\{user\_self\_id\})} provided the userId for C-API1. P-API1 returns the user's own userId, so C-API1 can only use the user's own userId.

	\textbf{BOPLA(CVE-2023-32310).} the attacker use API2\emph{(GET /api/share/treeList)} can view the panelId of the panels shared by other users and modify the panelId parameter of C-API2\emph{(POST /share/removePanelShares/\{panelId\})} to remove the panels shared by other users. However, under the protection of \tool, attackers do not have permission to use panelids shared by other users in C-API2. PanelId authorization fails because there is no data flow in the system for API2 to propagate panelId to C-API2. P-API2\emph{(POST /share/shareOut)} is associated with C-API2, and P-API2 isolates the panelId of C-API2 into the paneId of the panel that the user shares.
	
	\textbf{UASBF(CVE-2023-32310).} The msgId injection point of C-API3\emph{(POST /sys\_msg/batchDelete Body:msgId)} is a numeric type, and the attacker constantly guesses msgId of other users. \tool learns from the data stream that P-API3 provides msgIds for C-API3. P-API3\emph{(POST /sys\_msg/list)} generates the system messages users receive, so C-API3 can only delete messages using the msgId of the messages it receives.
	
	\textbf{Detection in the wild.} We use \tool to scan projects in Table \ref{table3} to obtain the MSG interval of the C-API resource ID. By passing the resource ID outside the permission into the C-API parameter, we determine whether there is a vulnerability according to the response result. \tool reports a total of 36 vulnerabilities, of which 35 vulnerabilities have been manually confirmed as real vulnerabilities, as shown in Table \ref{table8}. The false positive is caused by the modification of the resource ID during the propagation process. We have submitted vulnerability information to the CVE vulnerability database. Due to ethical considerations, we will disclose the vulnerability after notifying the manufacturer to repair it. The details of \tool report vulnerabilities are shown in Table \ref{table12}. 

     We also found an implementation-defective API (API1: \emph{findartbyuserid/\{userId\}}) whose function is to obtain a self-published article. The \emph{userId} parameter in API1 is the user's own userId, but the caller can obtain the article published by other users by modifying the \emph{userId}. API1 does not perform access control checks on \emph{userId}. Although there is another API in the system that can view all user-published articles, developers still need to add access control checks to API1 during implementation.

    \begin{table}[htbp]
    \setlength\tabcolsep{4.5pt}
	\centering
	\caption{BOLA vulnerabilities reported by \tool \label{table8}}
	\begin{tabular}{|l|l|l|l|l|l|l|l|l|l|l|}
		\hline
		\multirow{2}{*}{\textbf{Project}} & \multicolumn{2}{c|}{\textbf{SELECT}} & \multicolumn{2}{c|}{\textbf{INSERT}}  & \multicolumn{2}{c|}{\textbf{UPDATE}} & \multicolumn{2}{c|}{\textbf{DELETE}}& \multicolumn{2}{c|}{\textbf{Total}}\\ \cline{2-11} 
		& TP  & FP  & TP & FP & TP & FP& TP & FP& TP & FP\\ \hline
		\multicolumn{1}{|l|}{Blog} & 0  & 0 & 0 & 0  & 0 & 0& 0 & 0& 0 & 0 \\ \hline
        \multicolumn{1}{|l|}{BookStore} & 2  & 0 & 3 & 0  & 3 & 0& 3 & 0& 11 & 0 \\ \hline
        \multicolumn{1}{|l|}{Mall} & 1  & 0 & 0 & 0  & 1 & 0& 0 & 0& 2 & 0 \\ \hline
        \multicolumn{1}{|l|}{NewbellMall} & 0  & 0 & 0 & 0  & 0 & 0& 0 & 0& 0 & 0 \\ \hline
        \multicolumn{1}{|l|}{IceCms} & 1  & 0 & 3 & 0  & 0 & 0& 0 & 0& 4 & 0 \\ \hline
        \multicolumn{1}{|l|}{MusicwWbsite} & 3  & 0 & 4 & 0  & 3 & 0& 3 & 0& 13 & 0 \\ \hline
        \multicolumn{1}{|l|}{OnlineExam} & 1  & 0 & 1 & 1  & 1 & 0& 0 & 0& 3 & 1 \\ \hline
        \multicolumn{1}{|l|}{UniversityForum} & 0  & 0 & 2 & 0  & 0 & 0& 0 & 0& 2 & 0 \\ \hline
        \multicolumn{1}{|l|}{InformationSystem} & 0  & 0 & 0 & 0  & 0 & 0& 0 & 0& 0 & 0 \\ \hline
        \multicolumn{1}{|l|}{OnlineMall} & 0  & 0 & 0 & 0  & 0 & 0& 0 & 0& 0 & 0 \\ \hline
        \multicolumn{1}{|l|}{Total} & 8  & 0 & 13 & 1  & 8 & 0& 6 & 0& 35 & 1 \\ \hline
	\end{tabular}
    \end{table}
      \begin{table}[htbp]
    \setlength\tabcolsep{3pt}
		\small
		\caption{Identified unpublished vulnerabilities in projects\label{table12}}
		\centering
		\begin{tabular}{|c|c|l|}
			\hline
			\textbf{Project} &\textbf{Type}& \textbf{Description} \\
			\hline
			\makecell[c]{Mall}&\makecell[c]{SELECT}& Attackers get order informations of other users by modifying the orderId. \\
            \hline
            \makecell[c]{Mall}&\makecell[c]{UPDATE}& Attackers update order informations of other users by modifying the orderId. \\
			\hline
             \makecell[c]{Musicwebsite}&\makecell[c]{UPDATE} &Attackers update info of other users by modifying the userId. \\
             \hline
             \makecell[c]{Musicwebsite}&\makecell[c]{UPDATE} &Attackers update password of other users by modifying the userId. \\
			\hline
              \makecell[c]{Musicwebsite}&\makecell[c]{UPDATE} &Attackers update avatar of other users by modifying the userId. \\
            \hline
              \makecell[c]{Musicwebsite}&\makecell[c]{SELECT} &Attackers get collection's detail of other users by modifying the userId. \\
			\hline
              \makecell[c]{Musicwebsite}&\makecell[c]{INSERT} &Attackers add collection's of other users by modifying the userId. \\
			\hline
            \makecell[c]{Musicwebsite}&\makecell[c]{DELETE} & Attackers delete collection's of other users by modifying the userId. \\
			\hline
            \makecell[c]{Musicwebsite}&\makecell[c]{SELECT} & Attackers get collection's status of other users by modifying the userId. \\
			\hline
            \makecell[c]{Musicwebsite}&\makecell[c]{SELECT} & Attackers get rank of other users by modifying the userId. \\
			\hline
            \makecell[c]{Musicwebsite}&\makecell[c]{INSERT} & Attackers add rank of other users by modifying the userId. \\
			\hline
            \makecell[c]{Musicwebsite}&\makecell[c]{INSERT} & Attackers add comments of other users by modifying the userId. \\
			\hline
            \makecell[c]{Musicwebsite}&\makecell[c]{DELETE} & Attackers delete comments of other users by modifying the commentId. \\
			\hline
            \makecell[c]{Musicwebsite}&\makecell[c]{INSERT} & Attackers add support of other users by modifying the commentId and userId. \\
			\hline
            \makecell[c]{Musicwebsite}&\makecell[c]{DELETE} & Attackers delete support of other users by modifying the commentId and userId. \\
			\hline
            \makecell[c]{IceCMS}&\makecell[c]{SELECT} & Attackers get other users informations by modifying the userId. \\
			\hline
            \makecell[c]{IceCMS}&\makecell[c]{INSERT} & Attackers insert other users article comment by modifying the userId. \\
			\hline
            \makecell[c]{IceCMS}&\makecell[c]{INSERT} & Attackers insert other users square comment by modifying the userId. \\
			\hline
            \makecell[c]{IceCMS}&\makecell[c]{INSERT} & Attackers insert other users resource comment by modifying the userId. \\
			\hline
            \makecell[c]{BookStore}&\makecell[c]{DELETE} & Attackers delete addresses of other users by modifying the addressId. \\
			\hline
            \makecell[c]{BookStore}&\makecell[c]{UPDATE} & Attackers update addresses of other users by modifying the addressId. \\
			\hline
            \makecell[c]{BookStore}&\makecell[c]{INSERT} & Attackers add cart of other users by modifying the account. \\
			\hline
            \makecell[c]{BookStore}&\makecell[c]{DELETE} & Attackers delete cart of other users by modifying the account. \\
			\hline
            \makecell[c]{BookStore}&\makecell[c]{DELETE} & Attackers batch delete cart of other users by modifying the account. \\
			\hline  
            \makecell[c]{BookStore}&\makecell[c]{UPDATE} & Attackers update cart of other users by modifying the account. \\
			\hline 
            \makecell[c]{BookStore}&\makecell[c]{SELECT} & Attackers get cart of other users by modifying the account. \\
            \hline 
            \makecell[c]{BookStore}&\makecell[c]{INSERT} & Attackers init order of other users by modifying the account. \\
			\hline  
            \makecell[c]{BookStore}&\makecell[c]{INSERT} & Attackers add order of other users by modifying the account. \\
			\hline 
            \makecell[c]{BookStore}&\makecell[c]{SELECT} & Attackers get order of other users by modifying the account. \\
			\hline 
            \makecell[c]{BookStore}&\makecell[c]{UPDATE} & Attackers update order status of other users by modifying the orderId. \\
			\hline 
            \makecell[c]{OnlineExam}&\makecell[c]{UPDATE} & Attackers update password of other users by modifying the studentId. \\
			\hline 
           \makecell[c]{OnlineExam}&\makecell[c]{INSERT} & Attackers add score of other users by modifying the studentId. \\
			\hline 
            \makecell[c]{OnlineExam}&\makecell[c]{SELECT} & Attackers get score of other users by modifying the studentId. \\
			\hline 
            \makecell[c]{University Forum}&\makecell[c]{INSERT} & Attackers add post of other users by modifying the userId. \\
			\hline 
            \makecell[c]{University Forum}&\makecell[c]{INSERT} & Attackers add comment of other users by modifying the userId. \\
			\hline 
		\end{tabular}
	\end{table}

    \begin{tcolorbox}[colframe=black!50,colback=gray!10,left=1mm,right=1mm,top=0.5mm,bottom=0.5mm]
    {\textbf{Answer to RQ3:} \tool can defend against BOLA attacks in three attack modes and supports the detection of BOLA vulnerabilities in real-world projects.}
    \end{tcolorbox}

    \subsection{RQ4:How does \tool Compare with the SOTA Approach?} 
    \begin{table}[t!]
    \setlength\tabcolsep{4.5pt}
	\centering
	\caption{BOLA vulnerabilities reported by BOLARAY \label{table10}}
	\begin{tabular}{|c|l|l|l|l|l|l|l|l|l|l|}
		\hline
		\multirow{2}{*}{\textbf{Project}} & \multicolumn{2}{c|}{\textbf{SELECT}} & \multicolumn{2}{c|}{\textbf{INSERT}}  & \multicolumn{2}{c|}{\textbf{UPDATE}} & \multicolumn{2}{c|}{\textbf{DELETE}}& \multicolumn{2}{c|}{Total}\\ \cline{2-11} 
		& TP  & FP  & TP & FP & TP & FP& TP & FP& TP & FP\\ \hline
		\multicolumn{1}{|l|}{Blog} & 0  & 0 & 0 & 0  & 0 & 0& 0 & 0& 0 & 0 \\ \hline
        \multicolumn{1}{|l|}{BookStore} & 0  & 0 & 3 & 0  & 3 & 0& 3 & 0& 9 & 0 \\ \hline
        \multicolumn{1}{|l|}{Mall} & 0  & 0 & 0 & 0  & 1 & 0& 0 & 0& 1 & 0 \\ \hline
        \multicolumn{1}{|l|}{NewbellMall} & 0  & 0 & 0 & 0  & 0 & 0& 0 & 0& 0 & 0 \\ \hline
        \multicolumn{1}{|l|}{IceCms} & 0  & 0 & 3 & 0  & 0 & 0& 0 & 0& 3 & 0 \\ \hline
        \multicolumn{1}{|l|}{MusicwWbsite} & 0  & 0 & 4 & 0  & 3 & 1& 3 & 0& 10 & 1 \\ \hline
        \multicolumn{1}{|l|}{OnlineExam} & 0  & 0 & 1 & 0  & 1 & 1& 0 & 1& 2 & 2 \\ \hline
        \multicolumn{1}{|l|}{UniversityForum} & 0  & 0 & 2 & 0  & 0 & 0& 0 & 0& 2 & 0 \\ \hline
        \multicolumn{1}{|l|}{InformationSystem} & 0  & 0 & 0 & 0  & 0 & 0& 0 & 0& 0 & 0 \\ \hline
        \multicolumn{1}{|l|}{OnlineMall} & 0  & 0 & 0 & 0  & 0 & 0& 0 & 0& 0 & 0 \\ \hline
        \multicolumn{1}{|l|}{Total} & 0  & 0 & 13 & 0  & 8 & 2& 6 & 1& 27 & 3 \\ \hline
	\end{tabular}
    \end{table}
\subsubsection{Setup}
BOLARAY first analyzes the source code to infer the authorization mode of resources, and then verifies whether these models correctly implement access control checks. BOLARAY determines the access control mode of the resource based on the artificially summarized authorization mode. \tool does not need to classify resources into specific authorization models. Therefore, we focus on the difference between the authorization model of BOLARAY and the access control model of BOLAZ based on system logic. In order to eliminate the deviation caused by the authorization mode of BOLARAY inference resources, we first label the resource authorization model and use the API test method to verify the BOLA vulnerability.

BOLARAY needs to analyze \textit{admincolumn} to determine administrator permissions, but PHP and SpringBoot have different methods for determining administrator permissions. In some SpringBoot APIs, there is no administrator permission flag. \tool also detected APIs with non-administrator permissions. Therefore, in the comparative experiment, we remove the APIs of administrator permissions, registration and login, and reduced the false positives caused by BOLARAY due to \textit{admincolumn}.

\subsubsection{Results}
    The results of this comparison are summarized in Table \ref{table10}. BOLARAY correctly identified 27 true BOLA vulnerabilities, all of which were also disclosed by \tool. However, BOLARAY missed 8 vulnerabilities reported by \tool because it cannot identify BOLA vulnerabilities from SELECT statements. In contrast, \tool can analyze BOLA vulnerabilities with all types of SQL statements. BOLARAY reported 3 false vulnerabilities, but they are different from those reported by \tool, as shown in Table \ref{table11}.
\subsubsection{Case study of false positives} Through an in-depth investigation, we have summarized the causes of these false positives as follows.

   \textbf{Case 1: Application-level Authorization} \cite{huang_detecting_2024}. BOLARAY simplified the application-layer access control policy with the administrator role, and this simplification resulted in 2 false positives. In the online exam project, the teacher has the permission to modify, and delete all user data, but BOLARAY believes that non-administrator users can only operate on their own user data, resulting in false positives.

    \textbf{Case 2: Column-level Authorization} \cite{huang_detecting_2024}. Some applications require the authorization model to be more granular and limited to specific columns, leading to a false positive. In the music website project, the comment table is defined by BOLARAY as the ownership model, and a comment can only be updated/deleted by its owner. However, the comment-like API (/comment/like) modifies the comment's \emph{like\_count} column, which can be modified by all users.

    BOLARAY reported 2 more false vulnerabilities than \tool. This is because BOLARAY performs authorization checking based on the artificially summarized authorization model, which will lead to failure in some application scenarios. Its authorization model does not fit the workflow of the system, and wrongly determines the context between APIs. However, \tool defense rules based on the system's best-practice authorization logic solve these defects.
    
    \begin{table}[htbp]
    \setlength\tabcolsep{4.5pt}
		\caption{False positives reported by BOLARAY\label{table11}}
		\centering
		\begin{tabular}{|c|c|c|}
			\hline
			Project &API & Description\\
			\hline
            \makecell[c]{\multirow{2}*{\shortstack{OnlineExam}}} & \makecell[l]{PUT /student} & \makecell[l]{Application-level Authorization.}\\
            \cline{2-3} & \makecell[l]{DELETE /student\\/{studentId}} & \makecell[l]{Application-level Authorization.}\\
			\hline
			\makecell[c]{\multirow{1}*{\shortstack{MusicWebsite}}} & \makecell[l]{POST /comment/like} & \makecell[l]{Column-level Authorization.}\\
			\hline
		\end{tabular}
	\end{table}

    \begin{tcolorbox}[colframe=black!50,colback=gray!10,left=1mm,right=1mm,top=0.5mm,bottom=0.5mm]{\textbf{Insight \#2:} Combined with BOLARAY's authorization models, \tool can alleviate false negatives caused by ``client logic control (detailed in ~\ref{sec:MSG Rules Of Resource ID})".
    }\end{tcolorbox}
    \begin{tcolorbox}[colframe=black!50,colback=gray!10,left=1mm,right=1mm,top=0.5mm,bottom=0.5mm]{\textbf{Answer to RQ4:} Compared with BOLARAY, BOLAZ not only supports BOLA vulnerability detection for all types of SQL statements, but also breaks the limitations of the artificially summarized authorization model and has a lower false positive rate.
    }\end{tcolorbox}

\section{Discussion and Future Work}\label{sec:approach_limitations}
    \subsection{Threat to Validity.} 
    Through our experiments, we identified certain limitations in \tool's data flow analysis when the scenarios are relatively specific. First, as systems grow more complex in functionality and logic, it becomes challenging to interpret client-side control conditions using static code analysis. \tool sometimes was unable to determine which resource IDs produced by P-APIs are passed to C-APIs solely by analyzing condition code, particularly in scenarios involving client-controlled deletion and modification operations.

    Second, due to varying levels of developer expertise and coding practices, the same logic can be implemented in numerous ways. CodeQL cannot accommodate all coding patterns, leading to interruptions in some data flow analysis processes. As a result, API identification and association may fail, leaving some BOLA vulnerabilities unaddressed. In these cases, \tool will not restrict such APIs, and the vulnerabilities may persist.


    Third, \tool is currently implemented on the SpringBoot framework. A fact is that the propagation logic and behavior of resource IDs remains consistent in varied languages or frameworks. Hence, \tool is a generic approach at the methodological level.
 
    \subsection{Future Work.} 
    First, \tool currently only supports databases. In the future, we plan to investigate the potential of the MSG concept in other types of storage systems. Second, improvements are needed in \tool's front-end code analysis. We aim to integrate dynamic application security testing to better capture the logical relationships between P-APIs and C-APIs. Additionally, to address data flow interruptions in taint tracking, we will offer open programming interfaces, allowing users to adapt \tool to their system's coding patterns through customized rules.

\section{Related Work}\label{section:7}
\subsection{API Automated Testing} 
    Most automated tests for RESTful APIs focus on functional tests and are not used to detect BOLA vulnerabilities. However, we can draw ideas from API automated testing methods. 
	
	RestTestGen \cite{viglianisi_resttestgen_2020} analyzed the producers and consumers of resources in the API based on OAS. RESTler \cite{atlidakis_restler_2019} deduces producer-consumer dependencies of resources and resource IDs based on OAS. EvoMaster \cite{arcuri_restful_2019} generates test cases for RESTful APIs by analyzing the source code. However, EvoMaster believes that users can only operate on the resources they create, which does not apply to all application scenarios. Atlidakis et al. \cite{atlidakis_checking_2020} used automated API testing to verify whether the API violated the user-namespace rule. The rule holds that the resources produced by user A cannot be accessed by user B, and does not apply to all resource relationships. Corradini et al. \cite{corradini_automated_2023} Detect Mass Assignment Vulnerabilities in RESTful APIs using automated black-box testing. The method using common naming practices to identify the Resource ID is inaccurate. RESTest \cite{martin-lopez_restest_2020} is an automated black-box testing tool for RESTful APIs, which supports inference of dependencies between parameters.
	
	\subsection{Static Anlaysis-based BOLA Vulnerability Detection} 
    Application source code contains resource access patterns, so many tools use static analysis techniques to infer resource's authorization rules from the source code.

    SPACE \cite{near_finding_2016} requires developers to provide a mapping of application resources to the basic types that occur in SPACE catalog. SPACE uses symbolic execution to extract the data exposures from source code. Then SPACE checks whether each data exposure is allowed. Cancheck \cite{bocic_finding_2016} also requires developers to provide authorization rules for resources. RoleCast \cite{son_rolecast_2011} common software engineering patterns to determine resource authorization patterns, but RoleCast's patterns do not apply to all web applications. MACE \cite{monshizadeh_mace_2014} analyzes INSERT statements to determine relationships between users and resources, but MACE only supports the ownership model \cite{huang_detecting_2024} and cannot detect the BOLA vulnerability of SELECT statements. BOLARAY \cite{huang_detecting_2024} combines SQL and static analysis to automatically identify BOLA vulnerabilities in database-backed applications. However, BOLARAY also requires developers to mark DAL specifications, and does not support the detection of BOLA vulnerabilities in SELECT statements. The principle of MOCGuard \cite{liu_mocguard_2025} and BOLARAY to detect BOLA vulnerabilities is highly similar. They are based on human-summarized authorization models and use the relationship between the database tables corresponding to the resources to infer whether the user has permission to access the corresponding resources. The four authorization models summarized by BOLARAY are more comprehensive than MOCGuard and cover more scenarios. Similarly, MOCGuard also cannot handle the dynamic authorization logic of unknown systems due to the static nature of the artificially summarized authorization model.

    \subsection{Runtime Monitoring}
    There are also some works to enforce access control at system runtime. Nemesis \cite{dalton_nemesis_2009} ensures that only authenticated users can access resources under the manual authorization policy. FlowWatcher \cite{muthukumaran_flowwatcher_2015} found that the access control pattern of resources in most web applications is similar. Therefore, FlowWatcher requires developers to provide an access policy to detect whether the policy is violated during system runtime.

\section{Conclusions}\label{section:8}
	This paper proposes a defense framework for BOLA attacks, \tool, based on zero trust. The framework uses static taint tracking technology to obtain propagation data flows of resource IDs and determines APIs' role according to the production and consumption of resource IDs. To capture the context semantic relationship between P-APIs and C-APIs, \tool also tracks the propagation path of resource IDs between APIs. Experiments show that \tool can effectively identify and correlate APIs, low-performance overhead, and high security in multi-API vulnerability scenarios.

\section*{Acknowledgments}
This work is supported by the National Natural Science Foundation of China under Grant 62372323.

\bibliographystyle{ACM-Reference-Format}
\bibliography{ndss}

\end{document}